\documentclass[11pt,twoside,onecolumn]{article}
\usepackage[]{latexsym}
\usepackage{epsfig,epsf,epstopdf,graphicx}
\usepackage{amsmath,amssymb}
\RequirePackage[dvipsnames,usenames]{color}
\usepackage[titletoc,toc,title]{appendix}

\usepackage{amsfonts}
\usepackage{amsmath}
\usepackage{feynmf}

\usepackage{latexsym}
\usepackage{float} %Posiciona la imagen donde la pones el codigo exactamente.

\usepackage{slashed}
\usepackage{cite}
\usepackage[utf8]{inputenc} % acentos! Tildes en realidad
\usepackage{mathtools}
\usepackage[font=small,skip=-28pt]{caption}
\usepackage{color}
\DeclareFontFamily{OT1}{pzc}{}
\DeclareFontShape{OT1}{pzc}{m}{it}{<-> s * [1.10] pzcmi7t}{}
\DeclareMathAlphabet{\mathpzc}{OT1}{pzc}{m}{it}

\usepackage{commath} % Para módulo como \abs{}
\usepackage{accents} % Punto para la derivada \dt{} o \ddt{}
\usepackage{relsize} %agranda texto adentro de una ec. {\mathlarger{(x+y=z)}}
\usepackage[makeroom]{cancel} % Para tachar algo en una ecuacion
\usepackage{xcolor}
\usepackage[normalem]{ulem} %Permite subrayar varios renglones

\usepackage[]{hyperref} % Click in reference and goes there.

\newcommand{\ben}{\begin{equation}}
	\newcommand{\een}{\end{equation}}
\newcommand{\be}{\begin{equation*}}	
	\newcommand{\ee}{\end{equation*}}

\newcommand{\ba}{\begin{eqnarray}}
	\newcommand{\ea}{\end{eqnarray}}
\newcommand{\bal}{\begin{aligned}}
	\newcommand{\eal}{\end{aligned}}

\begin{document}

\title{Gravitational decoupled anisotropies in compact stars}

\author{Luciano Gabbanelli\\ 
	{\it Dept. de F\'isica Qu\`antica i Astrof\'isica, Institut de Ci\`encies del Cosmos (ICCUB),}\\
	{\it Universitat de Barcelona, Mart\'i ~i Franqu\`es 1, 08028 Barcelona, Spain.}\\
	\\
	{\'A}ngel Rinc{\'o}n and Carlos Rubio\\
	{\it Pontificia Universidad Cat\'olica de Chile,}\\
	{\it Av. Vicu\~na Mackenna 4860, Santiago, Chile.}
	\\[2ex]
	February, 2018. ICCUB-18-003
 	}

\date{}

\maketitle

\begin{abstract}
\noindent Simple generic extensions of isotropic Durgapal--Fuloria stars to the anisotropic domain are presented.
These anisotropic solutions are obtained by guided minimal deformations over a self gravitating isotropic system. 
When the isotropic and the anisotropic sector interacts in a purely gravitational manner, the conditions to decouple both sectors by means of the minimal geometric deformation approach are satisfied. Hence the anisotropic field equations are isolated resulting a more treatable set.
The simplicity of the equations allows one to manipulate the anisotropies that can be implemented in a systematic way to obtain different realistic models for anisotropic configurations. Later on, observational effects of such anisotropies when measuring the redshift are discussed.
To conclude, the application of the method over anisotropic solutions is generalized.
In this manner, different anisotropic sectors can be isolated of each other and modeled in a simple and systematic way. Besides, a generic property of the minimal geometric deformation approach, its noncommutativity, is discussed. This property duplicates the solutions obtained through this approach; anisotropies applied in the reversed order give different physically acceptable configurations.
\end{abstract}

%%%%%%%%%%%%%%%%%%%%%%%%%%%%%%%%%%%%%%%%%%%%%%%%%%%%%%
\section{Introduction}
%%%%%%%%%%%%%%%%%%%%%%%%%%%%%%%%%%%%%%%%%%%%%%%%%%%%%%
The study of analytical solutions of Einstein field equations plays a crucial role in the discovery and understanding of relativistic phenomena.
Theoretical grounds gives very few isotropic solutions under static and spherically symmetric assumptions. 
Worse yet, many less of these solutions have physical relevance passing elementary tests of astrophysical observations
\cite{Stephani,Delgaty,Semiz,Negi}.
Furthermore, no astronomical object is constituted of perfect fluids only; hence isotropic approximation
is likely to fail.

Anisotropic configurations have continuously attracted interest
and are still an active field of research. 
Strong evidence suggests that a variety of very interesting physical phenomena gives rise to a quite large number of local anisotropies, either for low or high density regimes (see \cite{Herrera} and references therein). For instance, among high density regimes, there are highly compact astrophysical objects with core densities ever higher than nuclear density $(\sim 3\times10^{17}\,\text{kg/m}^3)$ that may exhibit an anisotropic behaviour \cite{Ruderman}. Certainly, the nuclear interactions of these objects must be treated relativistically. The anisotropic behaviour is produced when the standard pressure is split 
in two different contributions: i) the radial pressure $p_r$ and ii) the transverse pressure $p_t$, which are not likely to coincide.

Anisotropies in fluid pressure usually arise due to the presence of a mixture of fluids of different types, rotation, viscosity, the existence of a solid core, the presence of a superfluid or a  magnetic field \cite{Mak:2001eb}. Even are produced by some kind of phase transitions or pion condensation among others \cite{Sokolov,Sawyer}. The sources of anisotropies have been widely studied in the literature, particularly for different highly compact astrophysical objects such as compact stars or black holes, either in 4 dimensions \cite{Maurya,Cho} as well as in the context of braneworld solution in higher dimensions \cite{Germani,OvalleCSinBW,OvalleSolidCrust}.

The main purpose of the present article is to generalize anisotropic analogous solutions of a particular kind of isotropic compact objects by means of the so-called {\it minimal geometric deformation} approach (MGD hereinafter) \cite{MGD-decoupling,Ovalle:2017wqi}. This method was originally proposed in the context of the Randall--Sundrum braneworld \cite{lisa1,lisa2} and was designed to deform the standard Schwarzschild solution \cite{MGDextended1,MGDextended2}. 
It describes the 4D geometry of a brane stellar distribution, hence obtaining braneworld corrections to standard GR solutions.
Therefore, it is a suitable method to obtain spherically symmetric and inhomogeneous stellar distributions that are physically admissible in the braneworld.
The key point of this approach is that the isotropic and anisotropic sectors can be split. Thus, the decoupling of both gravitational sources can be done in a simple and systematic way establishing a new window to search for new families of anisotropic solutions of Einstein field equations.

In this work we applied a gravitational decoupling through the MGD approach to derive exact and physically acceptable anisotropic interior solutions analogous to the 
Durgapal and Fuloria superdense star \cite{DF}. There have been several proposals of anisotropic models analogous to Durgapal--Fuloria compact stars \cite{MauryaDF,Ovalle:2008}; the MGD method seems to generalize them. The details of this method will be shown later, however the main lines goes as follows:
Let us start with a well known spherically symmetric gravitational source $T_{\mu\nu}^{\scriptscriptstyle (0)}$. This source can be as simple as one would desire; one can start with any known perfect fluid or even with vacuum itself. Any classical solution works as a seed for this method. After this, one switch on a new source of anisotropy
	\ben
	\widetilde{T}_{\mu\nu}=T_{\mu\nu}^{\scriptscriptstyle(0)}+\alpha\,T_{\mu\nu}^{\scriptscriptstyle(1)}\,.\een
When gravitational sources are couple via gravity only, i.e. they do not exchange energy-momentum among each other, the set of equations can be split into two contributions.
On the one hand, a well known sector is identified with the classical field equations of the chosen seed; the Durgapal--Fuloria solution for compact stars in our case. On the other hand, one is left with a simpler set of `pseudo-Einstein' equations for the sources of the anisotropy, to be solved. Combining both sectors a full anisotropic and physically consistent solution of Einstein field equations is obtained.
Of course one can switch on as many arbitrary sources of anisotropies $T_{\mu\nu}^{\scriptscriptstyle(i)}$ as desired, as long as a strategy to solve the new sector can be found.

This method for decoupling non-linear differential equations can be applied in a systematic way and has a vast unexplored territory where it could give different novel perspectives. MGD does not only give consistent interior solutions for different isotropic perfect fluid in GR; it could also be conveniently exploited in relevant theories such as $f(R)$--gravity \cite{fRgravity1,fRgravity2}, intrinsically anisotropic theories as Ho\v{r}ava--aether gravity \cite{Horava} or to study the stability of novel proposals of Black Holes, described by Bose Einstein gravitational condensate systems of gravitons \cite{BHasBEC:2018,Dvali:2012,Casadio:2016}. This is a robust method to extend physical solutions into an anisotropic domain preserving the physical acceptability.

The paper is organized as follows: after this introduction, we present the Einstein field equations for an anisotropic fluid. In Section \ref{MGD} we explain how the MGD approach is implemented to generate arbitrary anisotropic solutions. Section \ref{DFAnisotropic} is dedicated to apply this method to a particular seed, the Durgapal--Fuloria model for compact stars. In Section \ref{doubleAN} we extend the method to seeds which are already anisotropic. The last two sections are dedicated to discuss the main results and summarize our conclusions.

%%%%%%%%%%%%%%%%%%%%%%%%%%%%%%%%%%%%%%%%%%%%%%%%%%%%%%
\section{Anisotropic effective field equations}
%%%%%%%%%%%%%%%%%%%%%%%%%%%%%%%%%%%%%%%%%%%%%%%%%%%%%%
The simplest approach to describe compact distributions modelling stellar structures, is to restrict the metric to be static and spherically symmetric. In the usual Schwarzschild-like coordinates the line element takes the standard form
	\ben\label{MetricSchwStructure}
	\mathrm{d}s^2=e^{\nu}\,\mathrm{d}t^2-e^{\lambda}\,\mathrm{d}r^2-r^2\,(\mathrm{d}^2\theta+\sin^2\theta\,\mathrm{d}^2\phi)\,;\een
where the functions $\nu \equiv \nu(r)$ and $\lambda \equiv \lambda(r)$ depend on the radial coordinate only.
The encoded metric is a generic solution of the Einstein field equations 
	\begin{align}\label{EinEqFull}
	R_{\mu\nu}-\frac{1}{2}R\,g_{\mu\nu} =\kappa \widetilde{T}_{\mu\nu} \,, \end{align}
describing an anisotropic fluid sphere. The coupling constant between matter is given by $\kappa=8 \pi G/c^4$. Along these lines we will work in relativistic geometrized units, $G=c=1$. The observable features of the object will be determined by the exterior metric that will describe the geometry of the outer part. In the present article to maintain the treatment as simpler as possible, we will suppose a Schwarzschild vacuum outside.

The corresponding anisotropic effective stress-energy tensor $\widetilde{T}_{\mu\nu}$ is characterized by its diagonal components $\widetilde{\rho}$, $\widetilde{p}_r$ and $\,\widetilde{p}_t$, that are related to the geometric functions $\mu$, $\nu$ through \eqref{EinEqFull}. Explicitly,
	\begin{align}\label{EinEqGeneralT}
	\kappa \widetilde\rho &= \frac{1}{r^2}-e^{-\lambda}\left(\frac{1}{r^2}-\frac{\lambda'}{r}\right)\,,
	\\\label{EinEqGeneralR}
	-\kappa \widetilde{p}_r &= \frac{1}{r^2}-e^{-\lambda}\left(\frac{1}{r^2}+\frac{\nu'}{r}\right)\,,
	\\\label{EinEqGeneralA}
	-\kappa \widetilde{p}_t &=-\frac{1}{4}e^{-\lambda}\,\left(2\,\nu''+\nu'{}^2-\lambda'\,\nu'+2\,\frac{\nu'-\lambda'}{r}\right)\,.
	\end{align}
The prime stand for derivatives w.r.t. $r$. There is another equation consequence of the Bianchi identities: the covariant conservation of the stress-energy tensor
	\begin{equation}\label{ConsEqEffective}
	\nabla^\nu\,\widetilde{T}_{\mu\nu}=0\,.\end{equation}

Since the discovery of the first interior stellar solution by Schwarzschild \cite{Schwarzschild} and for several years,
stars interior were supposed to be constituted by perfect fluids. It was not until 1933 when Lema\^{i}tre \cite{Lemaitre} develop that spherically symmetry do not require the isotropic condition $\widetilde{p}_r=\widetilde{p}_t$, but only the equality of the two tangential pressures $\widetilde{p}_\theta=\widetilde{p}_\phi=\widetilde{p}_t$.
The system of equations \eqref{EinEqGeneralT}--\eqref{ConsEqEffective} governs the matter distribution within the star, which is assumed to be locally anisotropic (the radial and tangential pressure do not coincide). 
It is necessary to solve for five unknowns functions: two geometric functions, $\nu(r)$ and $\lambda(r)$; and three effective scalar functions, $\widetilde{\rho}(r)$, $\widetilde p_r(r)$ and $\widetilde p_t(r)$. However there are more unknowns than equations, hence the system is undetermined and constrains must be imposed. Some of them must be chosen by consistency of regularity, stability and (or) energy conditions of relativistic models; see for instance \cite{Herrera,Mak:2001eb,Ivanov:2017kyr}.

Throughout this article we will make use of the following representation for the effective energy-momentum tensor
	\ben\label{StressTensorEffective}
	\widetilde{T}_{\mu\nu}=T^{\scriptscriptstyle (PF)}_{\mu\nu}+\alpha\,\theta_{\mu\nu}\,.\een
The first term encodes a perfect fluid with isotropic pressure $p=p_r=p_t$,
	\ben\label{StressEnergyTensorPF}
	T^{\scriptscriptstyle (PF)}_{\mu\nu}=(\rho+p)\,u_\mu u_\nu-p\,g_{\mu\nu}\,.\een
$u_\mu$ is the normalized four-velocity field that accomplish $u_\mu u_\nu g^{\mu\nu}=1$. In our case, the perfect fluid will invariably be given by the Durgapal--Fuloria interior solution.
Under this representation, the anisotropic sector is described by the $\theta$--term. It describes additional gravitational sources responsible of the anisotropies. These source may contain new fields, whether scalar, vector or (and) tensor fields, coupled to gravity by means of a free dimensionless and constant parameter $\alpha$. One of the simplest and most treated examples in the literature are the anisotropies that may arise due to extra interactions resulting from the presence of charge \cite{MauryaDFQ}; besides there are plenty of complex treatments of anisotropies generated by other sophisticated physical fields \cite{Matt}.

The effective stress--energy tensor \eqref{StressTensorEffective} contributes at the level of Einstein equations with an effective energy density $\widetilde{\rho}$, an effective radial pressure $\widetilde p_r$ and an effective tangential pressure $\widetilde p_t$ defined as
	\begin{align}\label{AnEffDensity}
	  &\widetilde{\rho} =\rho+\alpha\,\theta_t{}^t\,,
	\\\label{AnEffPressureRadial}&\widetilde p_r = p-\alpha\,\theta_r{}^r\,,
	\\\label{AnEffPressureTan}&\widetilde p_t = p-\alpha\,\theta_\varphi{}^\varphi\,. \end{align} 
Thus, each magnitude is written as a deviation from the GR solution due to the presence of the $\theta$--term. 
The additive structure for the anisotropies allows the theory to have a straightforward limit to GR; setting $\alpha=0$ the standard Einstein equations for the perfect fluid are recovered.

Since the Einstein tensor is divergence free, under the representation taken in \eqref{StressTensorEffective} the covariant conservation equation \eqref{ConsEqEffective} yields
	\begin{align}\label{ConsEqExplicit}
	p'+\frac{\nu'}{2}\left(\rho+p\right)-\alpha\left[(\theta_r{}^r)'+\frac{\nu'}{2}(\theta_r{}^r-\theta_t{}^t)+\frac{2}{r}\left(\theta_r{}^r-\theta_\varphi{}^\varphi\right)\right]=0\,.\end{align}
This equation is a linear combination of \eqref{EinEqGeneralT} and \eqref{EinEqGeneralA}, as commonly happens in perfect fluid solutions of Einstein equations.
	
As this point let us remark the appearance of the anisotropy: there is not an \textit{a priori} restriction for the components of $\theta_{\mu\nu}$; however, if $\theta_r{}^r\neq\theta_\varphi{}^\varphi$ when solving the equation system \eqref{EinEqGeneralT}--\eqref{EinEqGeneralA}, we will be in the presence of the pressure anisotropy
	\ben\label{Anisotropy}
	\Pi\equiv\widetilde p_t-\widetilde p_r=\alpha\,\left(\theta_r{}^r-\theta_{\varphi}{}^{\varphi}\right).\een
Therefore, an isotropic stellar distribution (perfect fluid) becomes anisotropic when the $\theta$--term is turned on. 
In these lines we will follow a different approach to tackle the equation system \eqref{EinEqGeneralT}--\eqref{EinEqGeneralA}; we will address this system by means of the MGD method. This theory decouples the Einstein field equations when deforming the metric of the corresponding GR solution \cite{MGD-decoupling,Ovalle:2016pwp,Ovalle:2017khx,Ovalle:2017wqi}.

%%%%%%%%%%%%%%%%%%%%%%%%%%%%%%%%%%%%%%%%%%%%%%%%%%%%%%%%%%%%%%
\section{Minimal geometric deformation approach} \label{MGD}
\begin{figure}[b!]
	\centering\includegraphics[width=7.5cm]{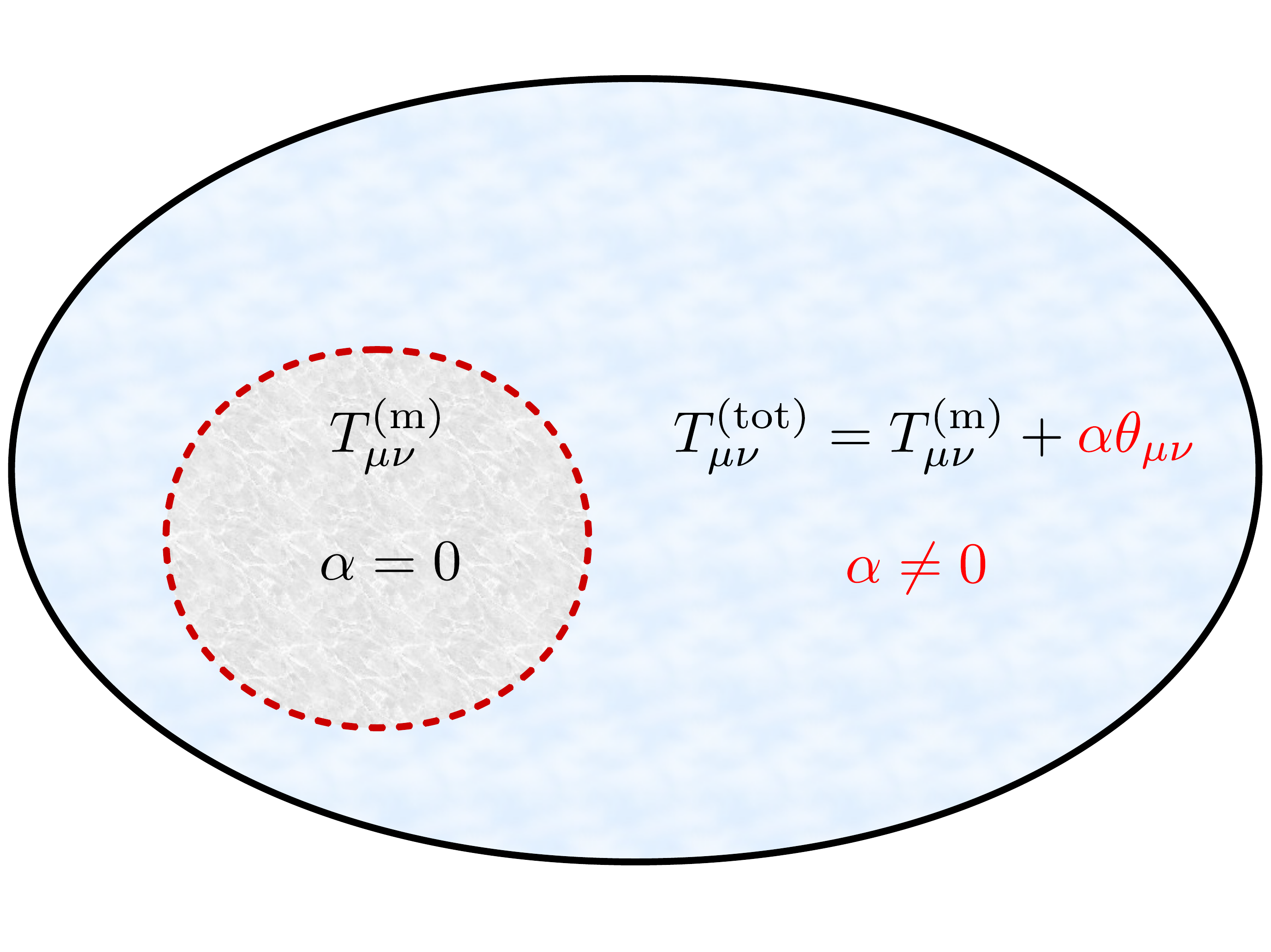}
	\vspace{20pt}\caption{\label{MGDTransf} 
	Any perfect fluid solution can be extended via the MGD approach \cite{Ovalle:2017wqi}. When $\alpha=0$ we are in the space of isotropic solution of Einstein equations, delimited by the red dashed line. This solution has a smooth extension to the anisotropic domain releasing the $\alpha$ to be nonnull.}\end{figure}

%%%%%%%%%%%%%%%%%%%%%%%%%%%%%%%%%%%%%%%%%%%%%%%%%%%%%%%%%%%%%%
With the aim of approaching the system of equations \eqref{EinEqGeneralT}--\eqref{EinEqGeneralA} in an alternative manner, a briefly review on the MGD procedure will be presented. This method produces anisotropic corrections to standard GR solutions providing physically
admissible non-uniform and spherically
symmetric stellar distributions.
The input (seed) is a known solution of Einstein equations: for instance the thermodynamic parameters satisfying \eqref{StressEnergyTensorPF}, and the corresponding geometric functions $\lambda(r)$ and $\nu(r)$. When a perfect fluid solution is taken as a seed, the isotropic condition $p_r=p_t=p$ is automatically accomplished. The method will produce a drift in the effective pressures such that $\widetilde{p}_r\neq\widetilde{p}_t$. For doing so, one implements the most generic {\it minimal geometric deformation} over the metric without breaking the spherical symmetry of the initial solution; this is
	\begin{align}
	&e^{+\nu(r)} \quad \rightarrow \quad e^{\nu(r)} + \alpha\,e^*(r)\,,	\\\label{MinGeoDefR}
	&e^{-\lambda(r)} \quad \rightarrow \quad \mu(r)  + \alpha\,f^*(r)\,,
	\end{align}
with $e$ and $f$ generic functions parametrizing the metric deformation.
In Figure \ref{MGDTransf} a schematic picture exemplifies how this method extends GR solutions to anisotropic domains when releasing $\alpha$. Even though the theory does not impose limits for the coupling strength, the physical acceptability of the new solution does so; if $\alpha$ is increased, the anisotropies become at some point unstable.

Although nothing prevent us from deforming the temporal component of the metric, it is general enough to start setting $e^*=0$; hence the effects of the $\theta_{\mu\nu}$ source undergo in a deformation over the radial coordinate only. The peculiarity of the MGD method is that it entails in its formulation a decoupling of the equations of motion. As a consequence of taking the $\theta$--sector as responsible of the {\it minimal distortion} of the metric, the system of equations \eqref{EinEqGeneralT}--\eqref{EinEqGeneralA} results quasi-decoupled: we obtain the Einstein equations for the chosen perfect fluid; and an effective `pseudo--Einstein' system of equations governing the $\theta$--sector. 
The only parameter that connects the two sectors is the temporal geometric function $\nu(r)$. At the same order as before, the temporal, radial and angular equations of motion relating the geometry of the spacetime to the thermodynamic characteristic of the perfect fluid sector reduce to
	\begin{align}\label{EinEqPFT}
	\kappa\,\rho&=\frac{1}{r^2}-\frac{\mu}{r^2}-\frac{\mu'}{r}\,,
	\\\label{EinEqPFR}
	-\kappa\,p&=\frac{1}{r^2}-\mu\,\left(\frac{1}{r^2}+\frac{\nu'}{r}\right)\,,
	\\\label{EinEqPFA}
	-\kappa\,p&=-\frac{1}{4}\left[\mu\,\left(2\,\nu''+\nu'{}^2+2\,\frac{\nu'}{r}\right)+\mu'\,\left(\nu'+\frac{2}{r}\right)\right]\,.\end{align}

The definition of a perfect fluid entails in itself the covariant conservation of the stress-energy tensor, i.e. 
	\ben\label{ConsEqPF}
	\nabla^\nu T^{\scriptscriptstyle (PF)}_{\mu\nu}=0\,.\een
The resulting equation is again a linear combination of the temporal and angular equations, \eqref{EinEqPFT} and \eqref{EinEqPFA}, and yields
	\ben\label{ConsEqExplicitPF}
	p'+\frac{1}{2}\nu'(\rho+p)=0\,.\een
It is worth noting that this system of equations is equivalent to Eqs. \eqref{EinEqGeneralT}--\eqref{EinEqGeneralA} if the coupling between the two sectors is set to zero; this is if the anisotropic sector vanishes.

The temporal component of the metric must satisfy binding conditions in the anisotropic sector: these are the remaining `pseudo--Einstein' field equations for the $\theta$--sector
	\begin{align}\label{PEFEAni0}
	\kappa\,\theta_t{}^t&=-\frac{f^*}{r^2}-\frac{f^*{}'}{r}\,,
	\\\label{PEFEAni1}
	\kappa\,\theta_r{}^r&=-f^*\,\left(\frac{1}{r^2}+\frac{\nu'}{r}\right)\,,
	\\\label{PEFEAni2}
	\kappa\,\theta_\varphi{}^\varphi&=-\frac{1}{4}\,\left[f^*\,\left(2\,\nu''+\nu'{}^2+\frac{2}{r}\,\nu'\right)+f^*{}'\,\left(\nu'+\frac{2}{r}\right)\right]\,.
	\end{align}
Once again, one has the corresponding conservation equation that is a consequence of \eqref{ConsEqEffective} and \eqref{ConsEqPF} satisfying separately. This equation is 
	\ben\label{ConsEqAnisotropy}
	\nabla^\nu\theta_{\mu\nu}=0\,,\een
and it is explicitly written as
	\ben\label{ConsEqExplicitTheta}
	(\theta_t{}^t)'-\frac{1}{2}\nu'(\theta_t{}^t-\theta_r{}^r)+\frac{2}{r}(\theta_r{}^r-\theta_\varphi{}^\varphi)=0\,.\een
This time the latter equation is not necessary linearly dependent of the `pseudo--Einstein' equations, and there is no reason why it should be. At this point it makes explicit that the interaction between the two sectors is purely gravitational; this is, from \eqref{ConsEqPF} and \eqref{ConsEqAnisotropy} is clear that each sector is separately conserved and there is no exchange of energy-momentum between them.

To conclude this section, let us summarize. First we started with an indeterminate system of equations \eqref{EinEqGeneralT}--\eqref{EinEqGeneralA}. Then, we performed a linear mapping of the radial geometric function of the metric \eqref{MinGeoDefR} that results in a `decoupling' of the Einstein field equations. We ended with two sets of equations: a perfect fluid sector $\{\rho\,;\,p\,;\,\nu\,;\,\mu\}$, given by \eqref{EinEqPFT}--\eqref{ConsEqExplicitPF} where everything is known once a perfect solution of GR is chosen; and a simpler sector of three linearly independent equations that can be chosen from \eqref{PEFEAni0} to \eqref{ConsEqExplicitTheta}, for determining four unknown functions $\{f^*\,;\,\theta_t{}^t\,;\,\theta_r{}^r\,;\,\theta_\varphi{}^\varphi\}$. Once the second sector is solved, we can identify directly the effective physical quantities introduced in \eqref{AnEffDensity}, \eqref{AnEffPressureRadial} and \eqref{AnEffPressureTan}.
At this point, is mandatory to recall that the underlying anisotropic effect which appears as a consequence of breaking the isotropic condition over the effective pressures, $\widetilde{p}_t \neq \widetilde{p}_r$, causes the appearance of the anisotropy $\Pi(\alpha;r)$ defined in Eq. \eqref{Anisotropy}.

%%%%%%%%%%%%%%%%%%%%%%%%%%%%%%%%%%%%%%%%%%%%%%%%%%%%%%
\section{Anisotropic Durgapal--Fuloria compact star} \label{DFAnisotropic}
%%%%%%%%%%%%%%%%%%%%%%%%%%%%%%%%%%%%%%%%%%%%%%%%%%%%%%
Let us proceed now to apply the MGD method with the aim of solving the Einstein field equations 
for the interior of anisotropic superdense stars. In the present work we will take as a seed the well-known Durgapal--Fuloria solution $\{\nu\,;\,\mu\,;\,\rho\,;\,p\}$ modeling compact stars. As explained before, once the MGD method is applied the system of equations \eqref{EinEqGeneralT}--\eqref{EinEqGeneralA} is decoupled. Half of the decoupled equations \eqref{EinEqPFT}--\eqref{EinEqPFA} are already solved once the relativistic perfect fluid is chosen. For instance, the thermodynamic functions that characterize the Durgapal--Fuloria solution are 
	\begin{align}
	\label{DFdensity}&\rho(r) = \frac{C\,(9+2\,Cr^2+C^2r^4)}{7\,\pi(1+Cr^2)^3}\,,
	\\\label{DFpressure}&p(r) = \frac{2\,C\,(2-7\,Cr^2-C^2r^4)}{7\,\pi(1+Cr^2)^3}\,,
	\end{align}
with $C$ an integration constant. The gravitational mass of a sphere of radius $r$ is obtained integrating the density inside the corresponding volume; in spherical coordinates is
	\ben\label{MassGR}
	m(r)=\int_{V}\rho\,\mathrm{d}V=\frac{4\,Cr^3(3+Cr^2)}{7(1+Cr^2)^2}\,.\een
This mass function has a well defined behaviour, vanishing at the center of the compact object, i.e. $m(r=0)=0$. It also determines the total mass evaluating the mass function at the surface, $m(r=R)=M$.

A massive object deforms the surrounding spacetime; the Durgapal--Fuloria solution is defined by the following metric components
	\begin{align}
	\label{PFSolgtt}
	&e^{\nu(r)}= A\,\left(1+Cr^2\right)^4\,,
	\\\label{PFSolgrr}
	&\mu(r) = 1 -\frac{2\,m}{r}\,.\end{align}
It is a custom in GR to write the radial component of the metric with the so-called {\it compactness parameter}, given by $\xi=2m/r$. The spacetime results regular everywhere, even at the center where $e^{\lambda(r=0)}\equiv\mu(r=0)=1$; $m$ vanishes faster than $r$ as one can easily check from \eqref{MassGR} inside \eqref{PFSolgrr}.
$A$ is the second (and last) integration constant to be determined using boundary conditions over the surface $r=R$. In the present article the outer metric will be chosen to satisfy the Schwarzschild form---for simplicity, an uncharged compact star.
Both constants $A$ and $C$ are positive; however they are expected to change as far as anisotropies begin to be considered.

The remaining equations after the decoupling, \eqref{PEFEAni0} to \eqref{PEFEAni2}, have to be solved if a generic anisotropic self-gravitating system is desired. The system of equation is as explain before underdetermined. A reasonable constrain is needed to close the system, but it is mandatory not to lose the physical acceptability of the solution. These issues will be discussed in what follows when three different anisotropic solutions (of many) are presented.

\subsection{Pressure--like constraint for the anisotropy}\label{DFAnisotropicMP}
%%%%%%%%%%%%%%%%%%%%%%%%%%%%%%%%%%%%%%%%%%%%%%%%%%%%%%
In order to close the system of equations \eqref{PEFEAni0}--\eqref{PEFEAni2}, additional information is needed. For instance, an equation of state for the source $\theta_{\mu\nu}$ or some physically motivated constrain on $f^*(r)$.
A first acceptable interior solution is deduced when forcing the associated radial pressure $\theta_r{}^r$ to mimic a physically acceptable pressure
	\ben\label{MimicConstrainP}
	\theta_r{}^r(r)=p(r)\,.\een
This means that one simple choice is to require that the stress-energy tensor for the perfect fluid coincides with the anisotropy in that direction.
As a consequence of \eqref{MimicConstrainP}, the radial Einstein equations for the GR solution \eqref{EinEqPFR} and the radial `pseudo-Einstein' equation \eqref{PEFEAni1} are equal. This gives immediately an expression for the radial component metric deformation
	\ben\label{DeformationMP}
	f^*(r)=-\mu+\frac{1}{1+r\,\nu'}\,.\een
The temporal component of the metric \eqref{PFSolgtt} remains non--deformed, so $\nu'$ is computed directly. The resulting deformed component, the radial one in \eqref{MinGeoDefR}, then becomes
	\ben\label{PFSolgrrAnisotropy}
	e^{-\lambda(r)}\quad\rightarrow\quad(1-\alpha)\,\mu+\alpha\,\frac{1+C r^2}{1+9\,Cr^2}\,.\een
It is explicit that when the $\alpha\rightarrow0$ limit is taken, one gets the non-perturbed Durgapal--Fuloria solution; particularly for the radial component of the metric,  $e^{-\lambda(r)}=\mu(r)$.

With the above considerations, the metric can be written in terms of an effective mass function  of the anisotropic sphere given by
 	\ben\label{MassFunctionMP}
	\widetilde{m}(r)=m-\alpha\,\frac{r\,f^*}{2}\,.\een
Expressed in this form, it is obtain one branch of MGD metrics that govern anisotropic interiors of GR solutions, whatever the GR solution is chosen. This branch corresponds to the pressure constrain imposed over the radial anisotropy. Therefore, the metric \eqref{MetricSchwStructure} is deformed to
	\ben\label{MetricMGD}
	\mathrm{d}s^2=e^{\nu}\,\mathrm{d}t^2-\left(1-\frac{2\,\widetilde m}{r}\right)^{-1}\,\mathrm{d}r^2-r^2\, \mathrm{d}\Omega^2\,.\een

As we have closed the system of equations with the constrain \eqref{MimicConstrainP}, we can compute all the effective magnitudes that characterized the fluid; but first, the values of the integration constants $A$ and $C$ are needed to be fixed.
This will be done by means of consistent matching conditions.

\subsubsection{Matching conditions}
A crucial aspect in the study of stellar distributions is the matching conditions at the star surface between the interior and the exterior spacetime geometries \cite{israel1,israel2}.
In our case, the interior stellar geometry is given by the MGD metric \eqref{MetricMGD}; while the outer part is assumed to be empty. Hence for $r\geq R$ the solution is given by the Schwarzschild vacuum solution. The continuity of the first fundamental form at the star surface $\Sigma$ (defined by $r=R$) is given by $\left[ds^2\right]_{\Sigma} = 0$.
This equation implies the continuity of the metric when crossing the surface and reads for the relevant components ($tt$ and $rr$--components) as
	\ben\label{FundamentalForm1}
	g_{tt}\,\bigr\rvert_{r=R^-}={g_{rr}}^{-1}\,\Bigr\rvert_{r=R^-}=1-\frac{2\,M_{\scriptscriptstyle Schw}}{R^+}\ .\een
The superindices stand for the region from where we approach the surface, either from inside with a minus sign, or from outside using the plus sign.
	
We must also take into account the Israel-Darmois matching condition at the stellar surface $\Sigma$ that gives the continuity of the second fundamental form $[G_{\mu\nu}\,x^\nu]_\Sigma=0$; $x^\nu$ is a unit vector. If we make use of the field equations \eqref{EinEqFull}, the continuity reads as $[\widetilde{T}_{\mu\nu}\,x^\nu]_\Sigma=0$. Using the full stress-energy tensor (\ref{StressTensorEffective}) and projecting in the radial direction $x^r=r$, is written as $[(T^{\scriptscriptstyle (PF)}_{rr}+\alpha\,\theta_{rr})\,r]_\Sigma=0$. This leads to
	\ben\label{FundamentalForm2Pressure}
	\widetilde p_r\hspace{4pt}\Bigr\rvert_{r=R^-}=\left(p-\alpha\,\theta_r{}^r\right)\Bigr\rvert_{r=R^-}=0\,;\een
where the effective pressure comes from Eq. \eqref{AnEffPressureRadial}. On the r.h.s. we are in vacuum, hence the pressure must nullify. 
The equation system has been closed with the constrain (\ref{MimicConstrainP}), therefore $\widetilde{p}(R)=0$ is equivalent to request $p(R)=0$ in (\ref{DFpressure}). This equivalence makes the constant $C$ not to vary from the perfect fluid solution once the anisotropies are considered. The value is
	\ben\label{AnConstantCMP}
	CR^2=\frac{-7+\sqrt{57}}{2}\,.\een

\begin{figure*}[b!]
	\centering
	\includegraphics[width=.45\linewidth]{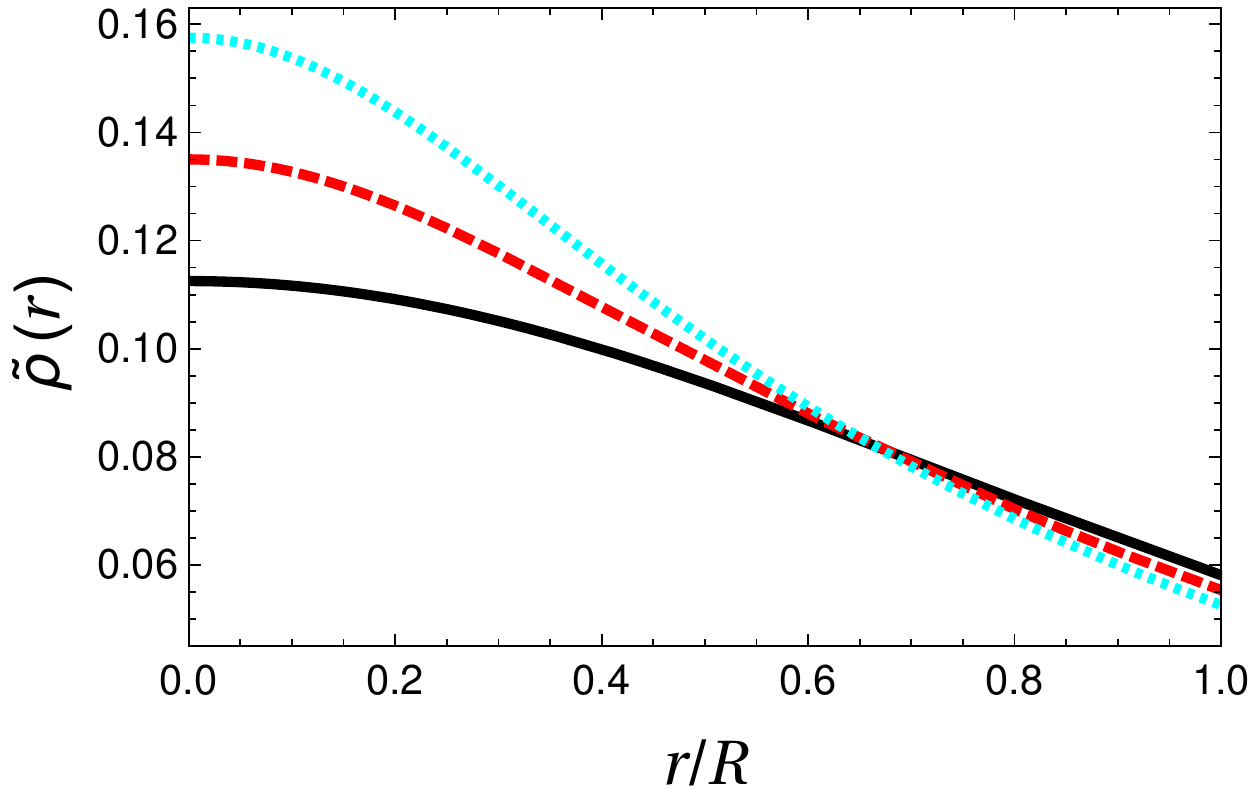}\quad
	\includegraphics[width=.45\linewidth]{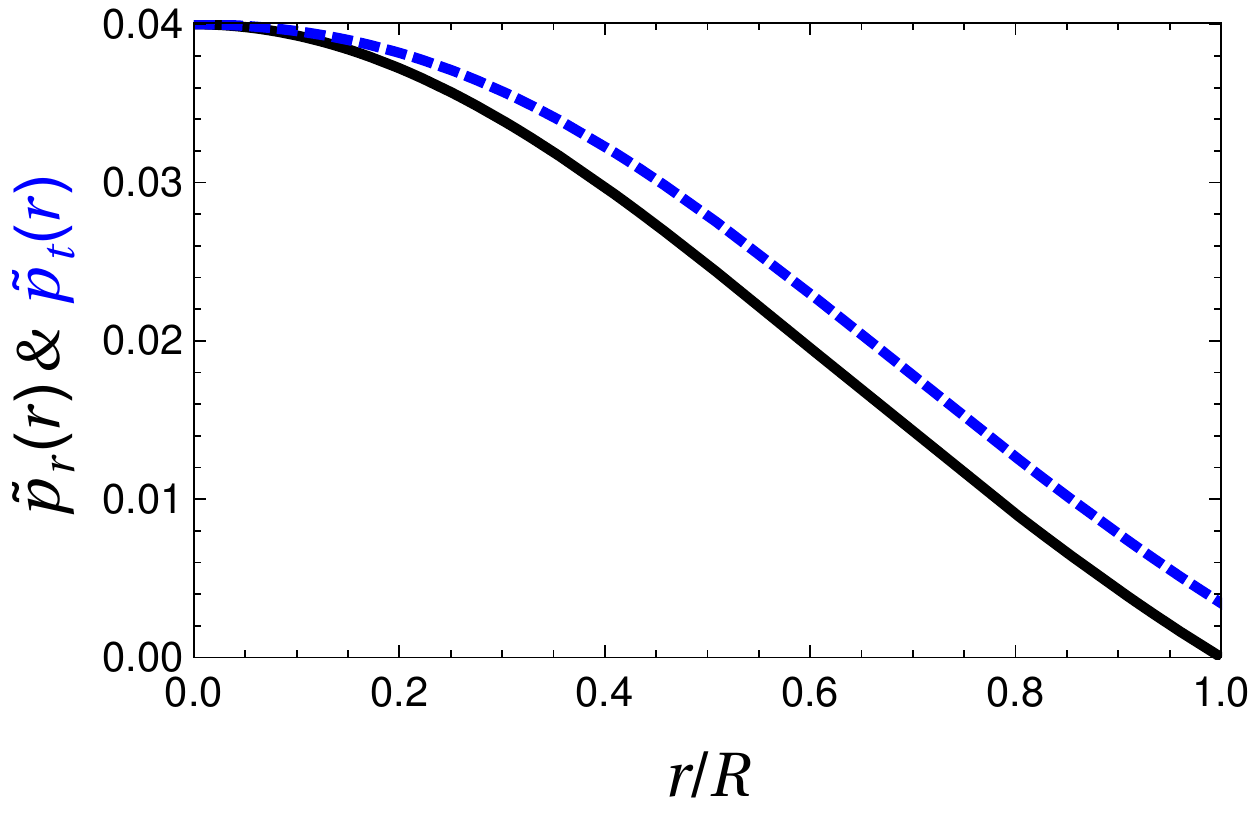}
	\medskip
	\includegraphics[width=.45\linewidth]{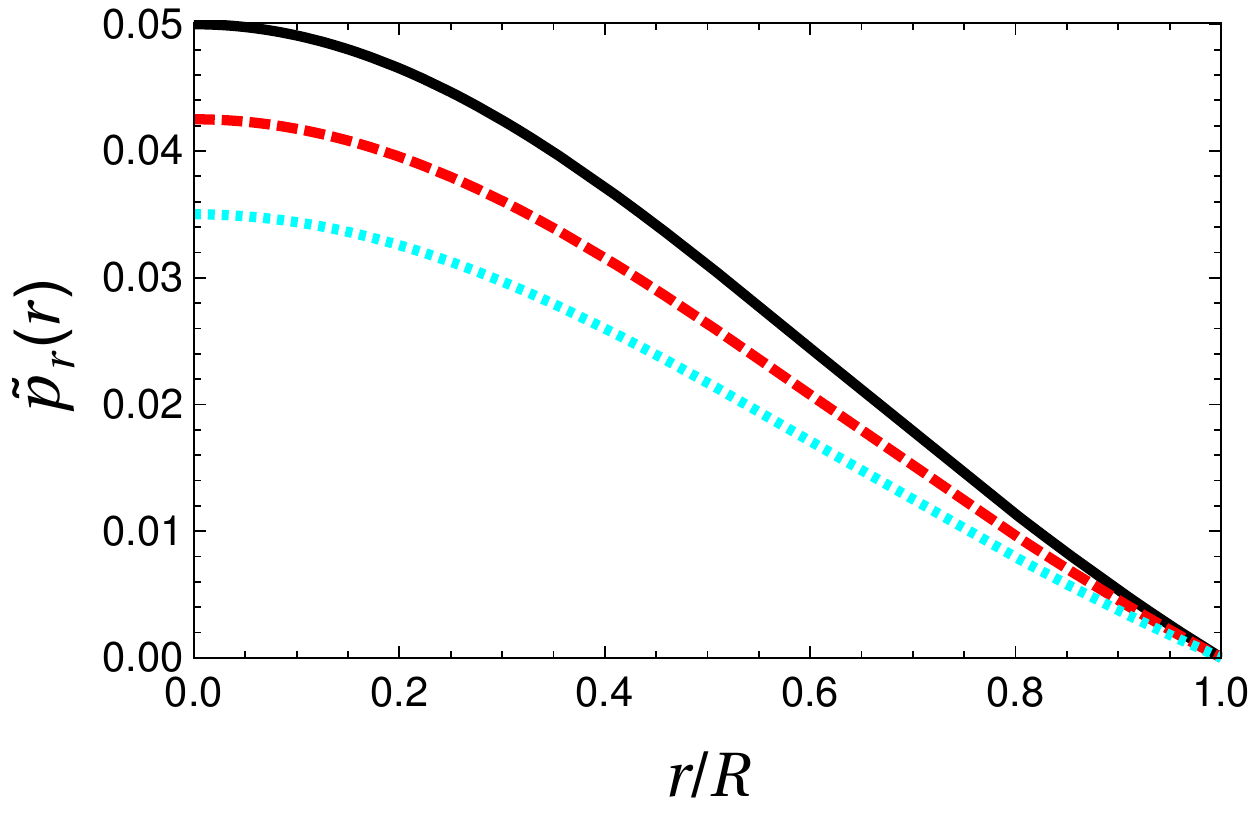}\quad
	\includegraphics[width=.45\linewidth]{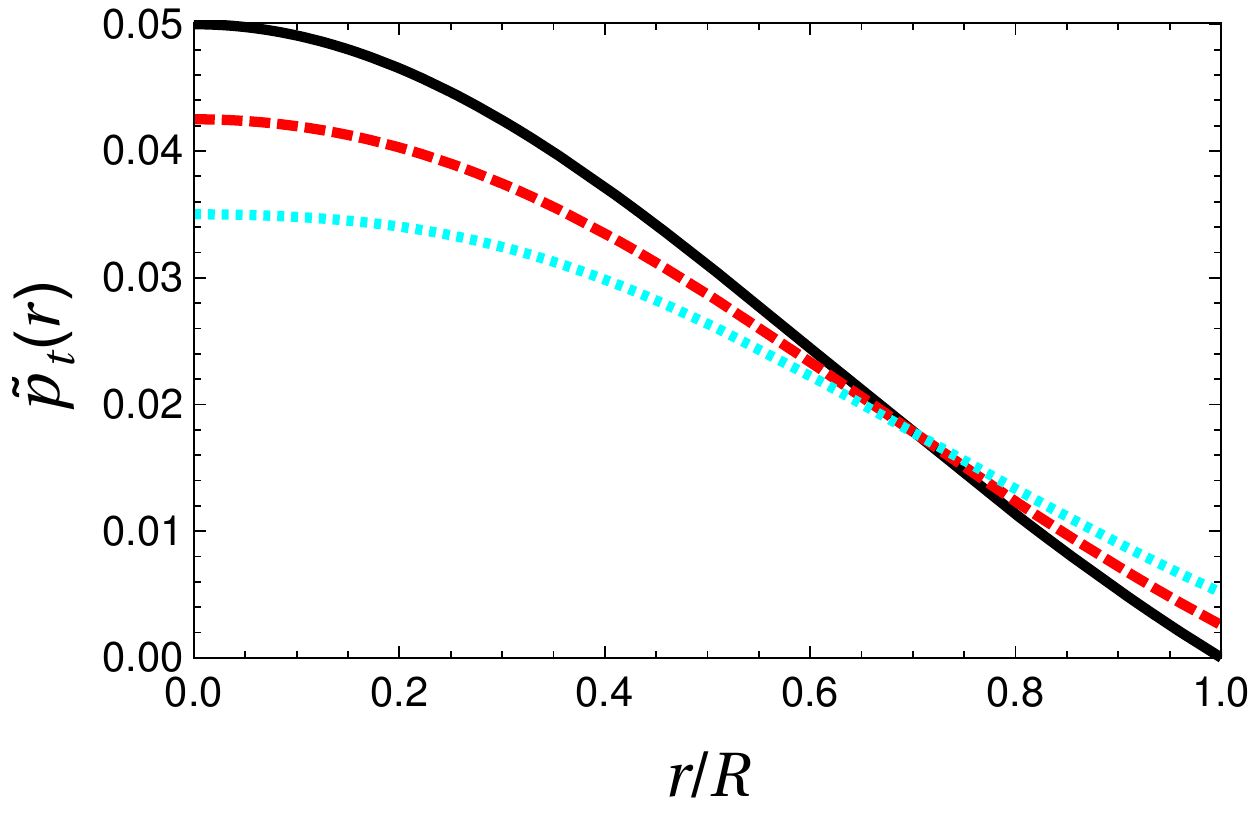}

	\vspace{25pt}
	\caption{Effective thermodynamic quantities for different values of $\alpha$ when the constrain over the anisotropy mimics a standard pressure $\theta_r\,^r=p$. The solid black line represents the standard Durgapal--Fuloria solution given by $\alpha=0$; $\alpha=0.15$ (dashed red line) and $\alpha=0.3$ (dotted cyan line) represent two anisotropic solutions. The second graph shows a comparison between the radial and tangential pressure for $\alpha=0.2$. The anisotropy causes the pressures values to drift apart.}\label{PlotMP}\end{figure*}

With the constant fixed, we have fully determined the effective radial pressure of the anisotropic Durgapal--Fuloria solution
	\ben\label{AnRadialPressureMP}
	\widetilde{p}_r(r;\alpha)=(1-\alpha)\,p\,.\een
A natural bound is obtained, $\alpha<1$. In Figure \ref{PlotMP}, it is shown the dependence of the pressure with a dimensionless radial coordinate $r/R$ for different values of $\alpha$.
At first sight one can observe that the higher $\alpha$ is, the smaller the radial pressure becomes. 
The decreasing of the radial pressure is needed to produce the pressure anisotropy that is reflected in a change over the tangential pressure along the surface. The expression for the later pressure is written as
	\ben\label{AnTanPressureMP} \widetilde{p}_t(r;\alpha)=\tilde{p}_r+\alpha\frac{6\,C^2r^2\left(1+3\,Cr^2\right)}{\pi\left(1+Cr^2\right)\left(1+9\,Cr^2\right)^2}\,.\een
The pressure (in both directions) must be a decreasing function along the radial coordinate. This condition restricts even more the values of $\alpha$; higher values immediately triggers instabilities. In light of what was written in \eqref{FundamentalForm2Pressure}, the tangential pressure \eqref{AnTanPressureMP} determines another physical constrain for $\alpha$: this pressure is meaningful as long as it remains positive everywhere $\widetilde{p}_t(r)>0$; hence, so must be $\alpha>0$ to not contradict this statement in the surface where $\widetilde{p}_r(R)=0$. From the latter equation, the anisotropy is directly computed; comparing with Eq. \eqref{Anisotropy}, we obtain
	\ben\label{AnAnisotropyMP}
	\Pi(r;\alpha)\equiv\alpha\,\frac{6\,C^2r^2\left(1+3\,Cr^2\right)}{\pi\left(1+Cr^2\right)\left(1+9\,Cr^2\right)^2}\,.\een

One can go on computing the remaining thermodynamic parameters. For instance the density can be expressed following \eqref{AnEffDensity} with the temporal component of the anisotropy given by \eqref{PEFEAni0}
	  \ben\label{AnDensityMP}
	  \widetilde{\rho}(r;\alpha)=\rho+\alpha\,\frac{2\,C\left(6-18\,Cr^2-257\,C^2r^4+15\,C^3r^6-9\,C^4r^8\right)}{7\pi\left(1+Cr^2\right)^3\left(1+9\,Cr^2\right)^2}\,.\een
Some comments are pertinent. The Durgapal--Fuloria solution is a fluid sphere with a solid crust. In Figure \ref{PlotMP} the density shows a discontinuity in the surface. The anisotropy smoothes this jump; the bigger the parameter $\alpha$ is, the lower the value of the density on the surface of the star. This behaviour immediately triggers the question on the profile of the effective mass function. This parameter has been defined in \eqref{MassFunctionMP} and together with \eqref{DeformationMP}, is written as
	\ben\label{AnMassMP}
	\widetilde{m}(r;\alpha) =\left[1+\alpha\,\frac{2\,(2-7\,Cr^2+C^2r^4)}{(3+Cr^2)(1+9\,Cr^2)}\right]m(r)\,.\een
An observer outside the star, would see a resulting mass $M_{\scriptscriptstyle Schw}$ surrounded by vacuum as it has been requested in (\ref{FundamentalForm1}). The continuity of the radial component of the metric (when crossing the star surface $\Sigma$) is direct: (\ref{MetricMGD}) identifies the Schwarzschild mass seen from outside with the effective mass of our solution ; i.e. $M_{\scriptscriptstyle Schw}\equiv\widetilde{m}(R)$.
Even more, a closer look at the mass function shows that the correction $(\alpha rf^*)/2$ in \eqref{MassFunctionMP} vanishes at the surface $r=R$. This means that the effective total mass of the star is the same as the isotropic total standard mass; $\widetilde{m}(R)\equiv m(R)$ from \eqref{MassGR}. This issue is not surprising at all. The anisotropy mimics the radial pressure, hence the radial and tangential pressure start to drift apart in the region close to the solid surface. For this anisotropic behaviour to happen, both pressures must decrease in magnitude at the inner region. Of course this pressure discrepancy with respect to the isotropic solution makes the density to be disturbed. The equilibrium between gravitational collapse and pressure repulsion is modified; hence the mass function is redistributed to the center of the star. Despite this, the total mass of the anisotropic object remains unmodified.

To conclude this section, let us determine the value of the remaining constant $A$ from \eqref{FundamentalForm1}. The temporal component of the MGD metric \eqref{PFSolgtt} should match smoothly with the outer Schwarzschild region
	\ben\label{FundamentalForm1gtt}
	g_{tt}\,\Bigr\rvert_{r=R^-}=A(1-C\,r^2)\,\Bigr\rvert_{r=R^-}=1-\frac{2\,M_{\scriptscriptstyle Schw}}{R^+}\,.\een
The constant $A$ remains unchanged. 
This constant close one branch ($\alpha$-dependent) of anisotropic solutions analogous to Durgapal--Fuloria; namely $\{\nu\,;\,\lambda\,;\,\widetilde\rho\,;\,\widetilde{p}_r\,;\,\widetilde{p}_t\}$. Of course, this solution is not unique. Different anisotropic solutions can be obtained starting from the Durgapal--Fuloria solution by means of requiring different constrains when closing the indeterminate system of equations. In next section we will consider a different constrain and we will see that a different anisotropic solution is obtained.

\subsection{Density--like constrain for the anisotropy}\label{DFAnisotropicMD}
%%%%%%%%%%%%%%%%%%%%%%%%%%%%%%%%%%%%%%%%%%%%%%%%%%%%%%%%%%%%%%
Another useful constrain that gives an acceptable physical solution, is to impose that the anisotropy `mimics' a density. This requirement is written as 
	\ben\label{MimicConstrainD}
	\theta_t{}^t(r)\equiv\rho(r)\een
and closes the system of equations (\ref{PEFEAni0})--(\ref{PEFEAni2}).
The consequence of this ansatz is direct, the temporal Einstein equation for the perfect fluid (\ref{EinEqPFT}) is identical to the temporal equation of motion for the $\theta_{\mu\nu}$--tensor (\ref{PEFEAni0}). Equaling both equations, one notes immediately that the resulting equation has a total derivative structure. The integration is straightforward
	\ben\label{EquationForg*} r(1-\mu+f^*)=K\quad\Longrightarrow\quad f^*(r)=\mu-1\een
where $K=0$ must be imposted for the invariants $R$, $R_{\mu\nu}R^{\mu\nu}$ and $R_{\mu\nu\gamma\sigma}R^{\mu\nu\gamma\sigma}$ to remain smooth and finite all over the inner region.
Note that with this constraint the radial deformation is again totally determined by the solution of the perfect fluid. Eventually, one computes the relevant component of the metric; the minimally deformed component is written as in \eqref{MinGeoDefR} (naming $\beta$ to the coupling between sectors) as
	\ben e^{-\lambda(r)}\quad\rightarrow\quad(1+\beta)\,\mu-\beta\quad\equiv\quad\mu-\beta\,\frac{8\,Cr^2(3+Cr^2)}{7\,(1+Cr^2)^2}\,.\een
We can write the metric with the structure used in \eqref{MetricMGD}. The effective mass is written as a minimal deviation from the GR mass $m(r)$ presented in \eqref{MassGR}
	\ben\label{AnMassMD}
	\widetilde{m}(r)=m+\beta\,\frac{\kappa}{2}\int \rho\,r^2 \mathrm{d}r = \left(1+\beta\right)m\,.\een
In the latter equality, if we make use of spherical coordinates and the corresponding relations, we have $\int\rho\,r^2\mathrm{d}r=m/\Omega_4$; where the 4--dimensional solid angle is $\Omega_4=\iint\mathrm{d}\Omega$ and $2\,\Omega_4/\kappa=1$. Of course, this is not surprising at all, the constrain for the anisotropy is to mimic the density, therefore the effective mass mimics the mass being proportional one to the other (unlike the previous case where the mass is exactly the same with respect to the standard GR solution).

Once the system is closed and the minimal deformation obtained, the remaining magnitudes are easily derived.
As before this will be done by means of the smooth matching between the inner and outer region of the star.

\subsubsection{Matching conditions}
Here we will reproduce the same steps that we have done before in order to find the constants $A$ and $C$; this time for the density ansatz \eqref{MimicConstrainD}. 
It is already known that the constant $C$ is determined by the second fundamental form \eqref{FundamentalForm2Pressure}. Its value is
	\ben\label{AnConstantCMD}
	CR^2=\frac{-7(1+2\,\beta)+\sqrt{(57+169\,\beta)(1+\beta)}}{2+9\,\beta}\,.\een
In this case, the anisotropic sector has an influence on the integration constant. It is explicitly seen that in the limit of no coupling $\beta\rightarrow0$ the constant from Durgapal--Fuloria is recovered.

\begin{figure*}[t!]
	\centering
	\includegraphics[width=.45\linewidth]{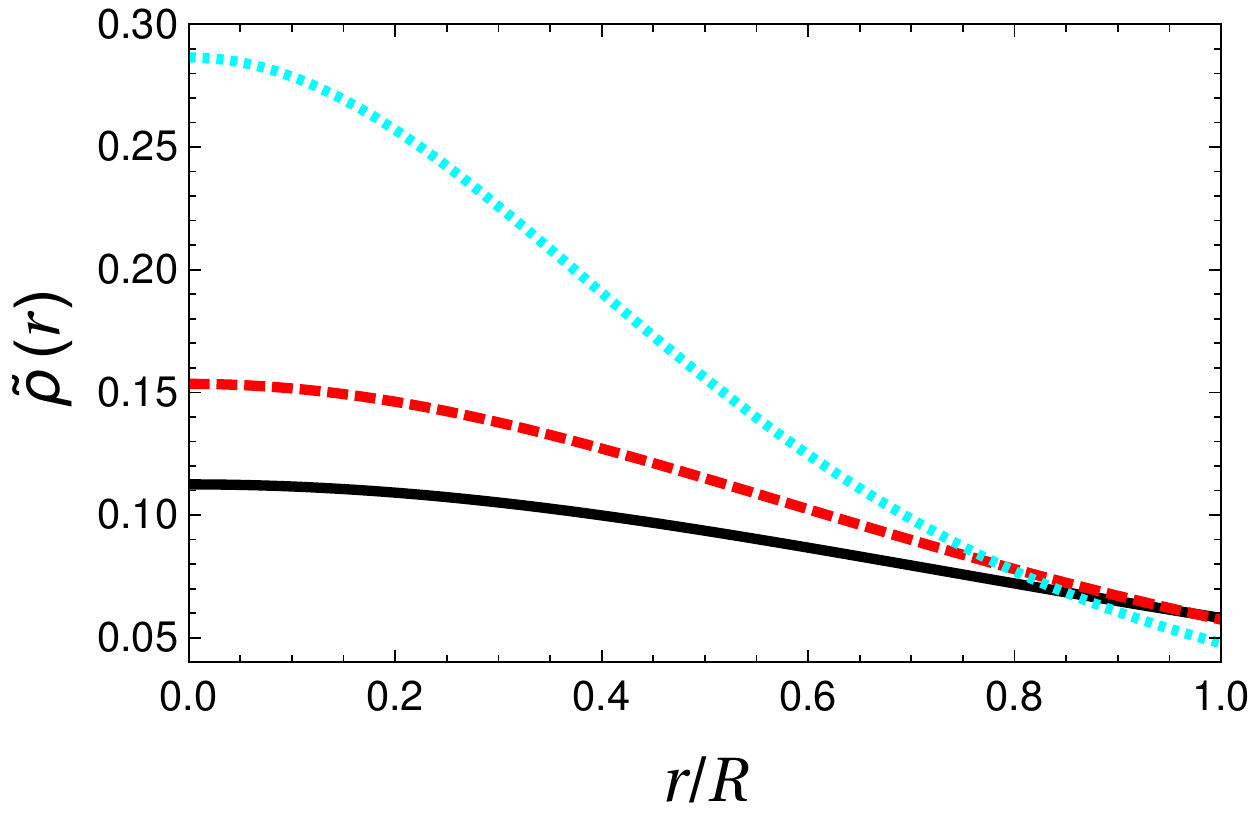}\quad
	\includegraphics[width=.45\linewidth]{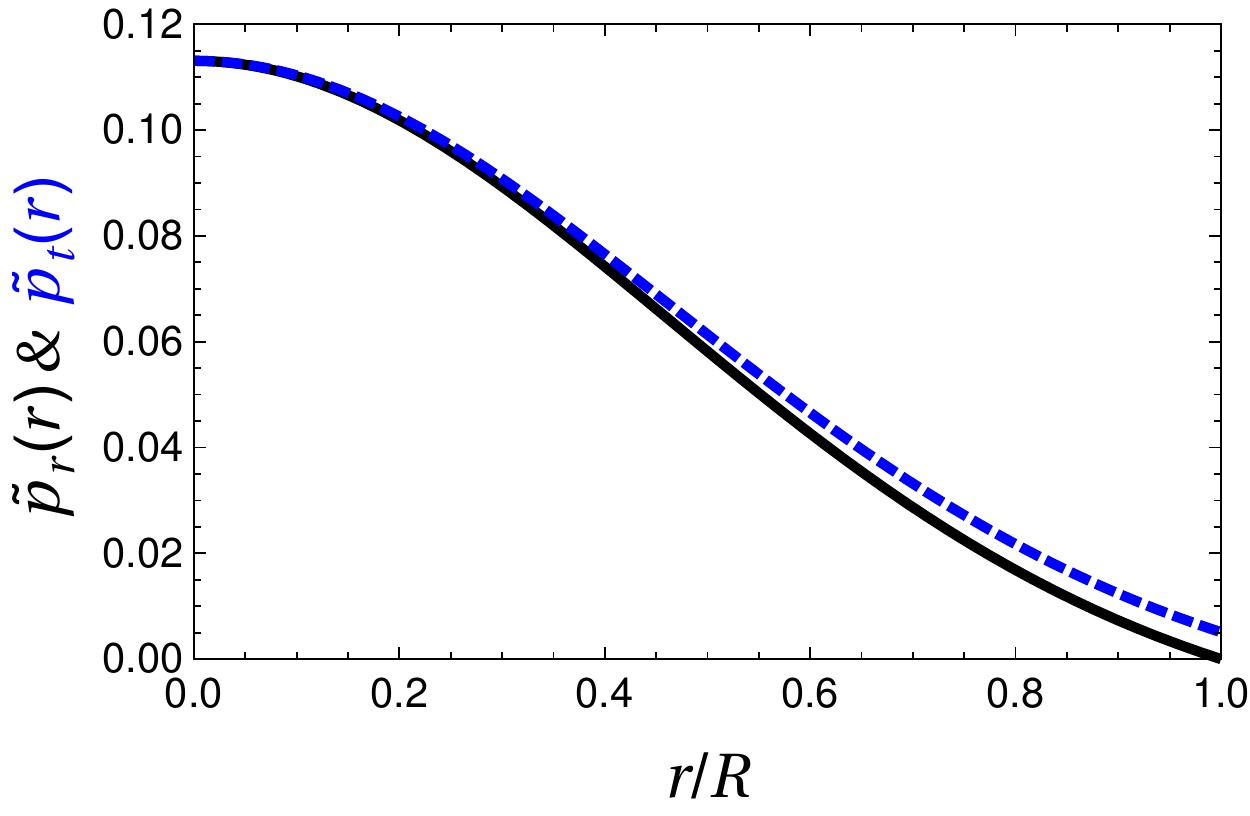}
	\medskip
	\includegraphics[width=.45\linewidth]{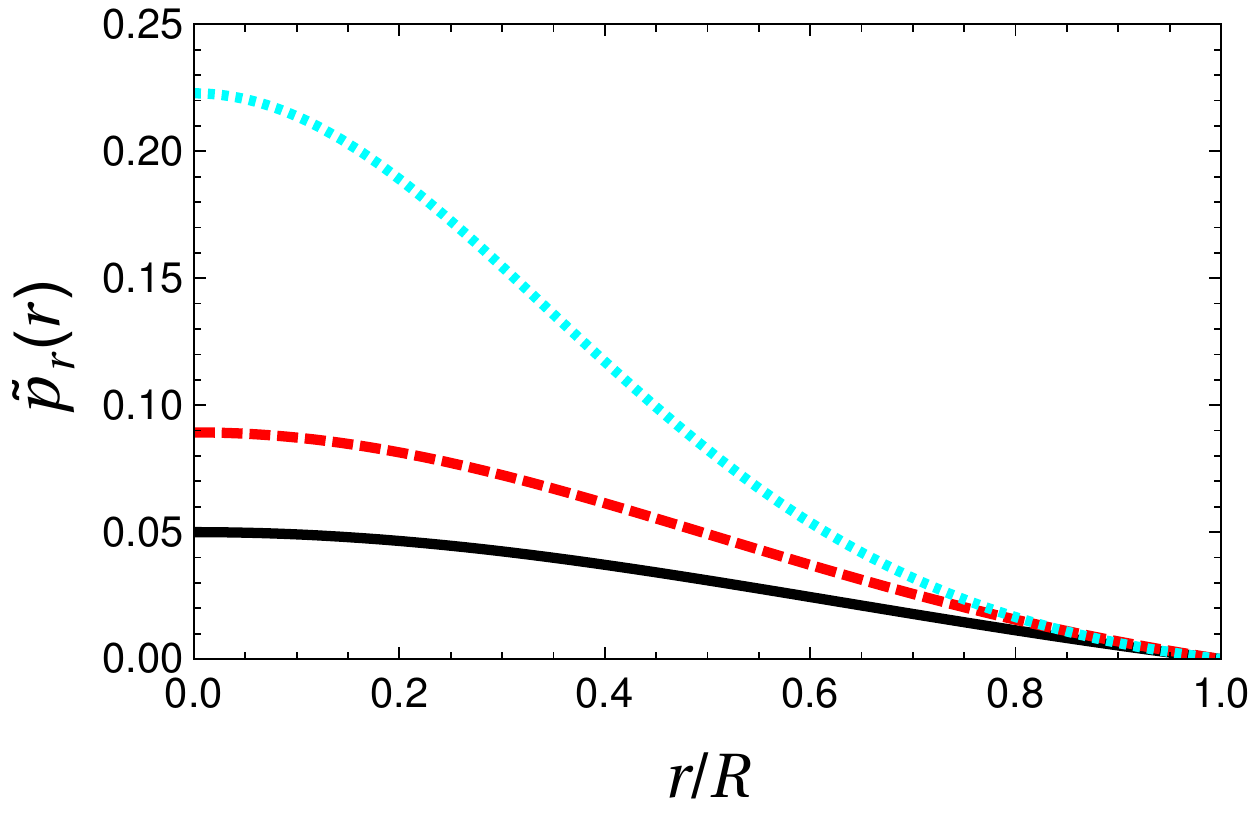}\quad
	\includegraphics[width=.45\linewidth]{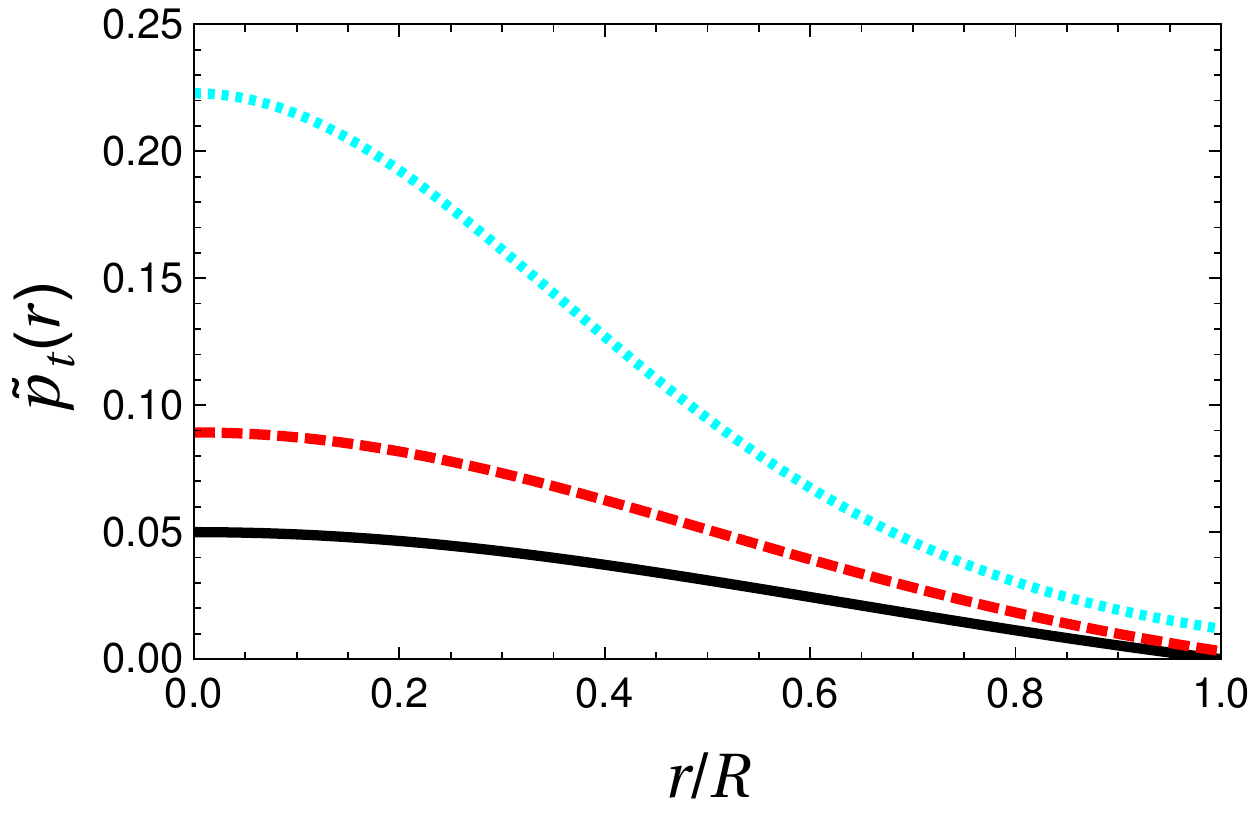}

	\vspace{25pt}
	\caption{Effective thermodynamic quantities for different values of $\beta$ when the anisotropy is forced to act as a density, $\theta_t\,^t=\rho$. The standard Durgapal--Fuloria solution has $\beta=0$ (solid black line); $\beta = -0.15$ (dashed red line) and $\beta = -0.3$ (dotted cyan line) represent two anisotropic solutions. The second set of curves shows the anisotropy over the pressure, $\widetilde{p}_r\neq\widetilde{p}_t$ for $\beta = -0.2$.\label{PlotMD}}\end{figure*}

Because of the ansatz where it has been required for the anisotropy to mimic the density, the effective value of the density is modified
	\ben\label{AnDensityMD}
	\widetilde{\rho}(r;\beta)=(1+\beta)\,\rho\,.\een
This is in complete accordance with the changes experienced by the mass. In what follows we will show that this solution only admits a {\it minimal geometric deformation} over the metric in only one `direction'; the `direction' to where the density and the mass is increase. The anisotropies restricted to the present constrain change the integration constants; for instance $C$ is $\beta$ dependent. An analysis over Eq. \eqref{AnConstantCMD} shows that $C$ increases when $\beta$ becomes more negative.
This behaviour makes Eqs. \eqref{AnMassMD} and \eqref{AnDensityMD} to increase when $\beta$ increase in modulus.
Of course, theses both parameters can not increase without a limit; as in the previous case, anisotropies develop instabilities. In the first curve of Figure \ref{PlotMD} it is seen how the density function increases in the inner region, while it slightly decreases its value over the surface's surroundings softening the crust. The mass function rises throughout the interior and the total effective mass is also increased.

The remaining thermodynamic parameters are the effective radial pressure
	\ben\label{AnRadialPressureMD}
	\widetilde{p}_r(r;\beta)=p-\beta\,\frac{C(3+Cr^2)(1+9\,Cr^2)}{7\pi(1+Cr^2)^3}\een
and the effective tangential pressure
	\ben\label{AnTanPressureMD}
	\widetilde{p}_t(r;\beta)=\widetilde{p}_r-\beta\,\frac{C^2r^2}{\pi(1+Cr^2)^2}\,.\een
As can be seen in the second graph of Figure \ref{PlotMD}, being the latter pressure different from the former, the anisotropy is developed. In the third and fourth set of curves, it is seen that both pressures are enhanced for the anisotropy to take place. When the mass increases, the enhancement of the density requires higher pressures for stability reasons.

In order to get some insight in the underlying sign for $\beta$ in the new solution, we will focus ourselves in the anisotropy, given by
	\ben\label{AnAnisotropyMD}
	\Pi(r;\beta)=-\beta\,\frac{C^2r^2}{\pi(1+Cr^2)^2}\,.\een
It is worth to note that this magnitude can not be negative (let us remind that $C>0$) because if this were so, over the surface of the star where $\widetilde{p}_r(R)=0$, we would have a negative tangential pressure $\widetilde{p}_t(R)<0$ which is not physically acceptable. Therefore, positive pressures implies negative values for $\beta$. Now we have a physical domain for $\beta$. The constant $A$ is found in an analogous manner than in the previous section, i.e. by means of Eq. \eqref{FundamentalForm1gtt}. The value of this constant changes with $\beta$. The usual constant of Durgapal--Fuloria is recovered in the limit of $\beta\rightarrow0$ as it should be.

\subsection{Detectability and observational differences in anisotropic distributions}
One of the many remarkable predictions of the theory of general relativity is the time dilation within a gravitational well. This results in footprints in the lines of the spectrum shifting towards the red. Although it is a useful quantity, particularly for astronomers, which allows to get some insight into compact stars physics, this effect is extremely difficult to deal with because of its complexity to be disentangle from the displacements and alterations due to various causes such as the Doppler, Zeeman and pressure effects among others. Theoretical derivations states that the redshift factor
associated to a star comes when relating the proper time $\tau$ of the object with the observer clock $t$. This relation is given by the standard formula $\mathrm{d}\tau^2=g_{tt}\,\mathrm{d}t^2$ that yields the following for the surface redshift
	\ben\label{Redshift}
	1+z=\frac{\nu_e}{\nu_o} =\frac{1}{\sqrt{g_{tt}(R)}}\,.\een
Therefore the relation between the emitted and observed frequency makes the redshift manifest \cite{Hladik}.

Over the years, the study of anisotropies in compact objects has received considerable attention and this parameter is a simple way to contrast theory with observations. Formula \eqref{Redshift} relates the measured redshift with the {\it compactness parameter} $\xi=2m/r$ of the star (that depends on the anisotropic coupling in this case $\alpha/\beta$) introduced after Eq. \eqref{PFSolgtt}. 
An observer outside would see the Schwarzschild metric, hence the redshift, that depends on $\xi$, is directly related with the total mass of the star (we are generating anisotropic contributions over fixed radius stars). 
Now, we want to investigate how this parameter evolve in our particular solutions.

Let us start with the fist solution derived in Section \ref{DFAnisotropicMP}. The interest should focus in the mass function \eqref{AnMassMP} evaluated over the surface. As we have explained after this equation the Schwarzschild mass remains unmodified with respect to the Durgapal--Fuloria mass, $M_{\scriptscriptstyle Schw}\equiv\widetilde{m}(R)=m^{\scriptscriptstyle (DF)}(R)$. Hence, there is no observational evidence to differentiate an isotropic star to these anisotropic counterpart.

On the contrary, things change in the second case. The solution obtained in Section \ref{DFAnisotropicMP} is based on an increment of the total mass. Then a shift occurs when observing this anisotropic configuration
	\ben\label{RedshiftMD}
	z(\beta)=\left[1-\frac{2M_{\scriptscriptstyle Schw}(\beta)}{R}\right]^{-1/2}-1\,.\een
The Schwarzschild mass coincides to the mass function over the surface; i.e. $M_{\scriptscriptstyle Schw}=\widetilde{m}(R)$ in Eq. \eqref{AnMassMD}. As it has been explained, the mass increases when $\beta$ increases in modulus. Therefore, the compactness parameter is also increased; the star becomes more and more denser with $\beta$. Then, the parenthesis in \eqref{RedshiftMD} decrease and $z(\beta)$ grows when $\abs{\beta}$ grows. This means that this anisotropic contributions increases the gravitational redshift as it is expected when the stars are more dense.

\section{Anisotropizing an anisotropic Durgapal--Fuloria star}\label{doubleAN}
In section \ref{MGD}, we present a method to generate different anisotropic solutions of Einstein field equations using any well known perfect fluid as a seed. After this, we apply this prescription to the Durgapal--Fuloria perfect sphere. In section \ref{DFAnisotropic}, with some reasonable constrains we found two novel physical anisotropic solutions analogous to the Durgapal--Fuloria compact star.

The decomposition of Einstein equations (\ref{EinEqGeneralT})--(\ref{EinEqGeneralA}) stem on the minimal geometric deformation (\ref{MinGeoDefR}); the anisotropic sector (\ref{PEFEAni0})--(\ref{PEFEAni2}) is decoupled with respect to any perfect fluid sector (\ref{EinEqPFT})--(\ref{EinEqPFA}). However, there is no need for the known sector to be a perfect fluid solution exclusively. Whatever solution of Einstein field equations, either a perfect or an anisotropic fluid, work as a seed for implementing the MGD decomposition. For instance, we can take any of the two previous founded solutions; e.g. the one obtained in section \ref{DFAnisotropicMP} given by
$\{\nu\,;\,\widetilde{\mu}\,;\,\widetilde{p}_r\,;\,\widetilde{p}_t\}$. So as not to obscure how the method works, we will {\it minimally deform} the anisotropic solution along the radial component of the metric. While the temporal geometric function in Eq. (\ref{PFSolgtt}) remains unchanged, the {\it }minimal distortion takes place only over the radial component
	\ben\label{MinGeoDefRMPMD}
	e^{-\widetilde\lambda(r)} \quad \rightarrow \quad e^{-\bar\lambda(r)}=\widetilde\mu(r)+\beta\,g^*(r)\,.\een
of an anisotropic metric solution of Einstein equation; in this case (\ref{MetricMGD}). This deformation is caused by new generic sources of anisotropies (called $\psi_{\mu\nu}$ in order to avoid confusion with the deformed seed by $\theta_{\mu\nu}$) that acts over the anisotropic energy momentum tensor \eqref{StressTensorEffective}
	\ben\label{StressTensorEffectiveMPMD}
	\overline{T}_{\mu\nu}=\widetilde{T}_{\mu\nu}+\beta\,\psi_{\mu\nu}\,.\een
The Einstein field equations connecting the latter effective stress-energy tensor to the spacetime curvature are
	\begin{align}\label{EinEqAniGeneralT}
	\kappa\,\bar{\rho}&=\frac{1}{r^2}-e^{-\bar\lambda}\left(\frac{1}{r^2}-\frac{\bar\lambda'}{r}\right)\,,
	\\
	\label{EinEqAniGeneralR}
	-\kappa\,\bar{p}_r&=\frac{1}{r^2}-e^{-\bar\lambda}\left(\frac{1}{r^2}+\frac{\nu'}{r}\right)\,,
	\\
	\label{EinEqAniGeneralA}
	-\kappa\,\bar{p}_t&=-\frac{1}{4}e^{-\bar\lambda}\,\left(2\,\nu''+\nu'{}^2-\bar\lambda'\,\nu'+2\,\frac{\nu'-\bar\lambda'}{r}\right)\,.\end{align}

The {\it minimal geometric deformation} \eqref{MinGeoDefRMPMD} decouples the two anisotropic sectors.
On the one hand, the seed sector characterized by the density given by (\ref{AnDensityMP}), the radial pressure $\widetilde{p}_r$ (\ref{AnRadialPressureMP}) and the tangential anisotropic pressure $\widetilde{p}_t$ obtained in (\ref{AnTanPressureMP}). This parameters solve the already known equation system
	\begin{align}\label{EinEqAniSeedT}
	\kappa\widetilde\rho&=\frac{1}{r^2}-\frac{\widetilde\mu}{r^2}-\frac{\widetilde\mu'}{r}\,,
	\\\label{EinEqAniSeedR}
	-\kappa\widetilde{p}_r&=\frac{1}{r^2}-\widetilde\mu\,\left(\frac{1}{r^2}+\frac{\nu'}{r}\right)\,,
	\\\label{EinEqAniSeedA}
	-\kappa\widetilde{p}_t&=-\frac{1}{4}\left[\widetilde\mu\,\left(2\,\nu''+\nu'{}^2+2\,\frac{\nu'}{r}\right)+\widetilde\mu'\,\left(\nu'+\frac{2}{r}\right)\right]\,;\end{align}
and on the other hand, we are left with the new anisotropic sector for $\psi_{\mu\nu}$ completely decoupled. In this sector we have the following `pseudo-Einstein' equations
	\begin{align}\label{EinEqPseudoAniT}
	\kappa\,\psi_t{}^t&=-\frac{g^*(r)}{r^2}-\frac{g^*{}'(r)}{r}\,,
	\\\label{EinEqPseudoAniR}
	\kappa\,\psi_r{}^r&=-g^*\,\left(\frac{1}{r^2}+\frac{\nu'}{r}\right)\,,
	\\\label{EinEqPseudoAniA}
	\kappa\,\psi_\varphi{}^\varphi&=-\frac{1}{4}\,\left[g^*\,\left(2\,\nu''+\nu'{}^2+\frac{2}{r}\,\nu'\right)+g^*{}'\,\left(\nu'+\frac{2}{r}\right)\right]\,.\end{align}

Subsequently, a constrain over the solution must be imposed; the system is indeterminate. Until now we used a constrain that mimics the pressure to obtain the seed solution; a density constrain will be applied now to combine both previously found solutions.
The ansatz then is to require
	\ben
	\psi_t{}^t\equiv\widetilde\rho(r)\,.\een
Now the steps that follow are known. Eqs. (\ref{EinEqAniSeedT}) and (\ref{EinEqPseudoAniT}) equals and give a total derivative equivalent to \eqref{EquationForg*}. The solution for the second minimal deformation function is straightforward because the constant of integration is again null
	\ben
	g^*=\widetilde{\mu}-1\,.\een
The radial metric component from \eqref{MinGeoDefRMPMD} then promotes to
	\ben
	e^{-\overline{\lambda}(r)}=(1-\alpha)(1+\beta)\mu+\,\frac{\alpha(1+\beta)}{1+r\,\nu'}-\beta\,;\een
where the anisotropic radial component of the metric \eqref{PFSolgrrAnisotropy} has been used.

The effective radial pressure $\bar{p}_r=\widetilde{p}_r-\beta{\psi_r}^r$ is computed from \eqref{EinEqAniGeneralR}
	\begin{align}
	\label{AnRadialPressureMPMF}
	\bar{p}_r(r;\alpha,\beta)&=\left[1-\alpha(1+\beta)\right]p-\beta\,\frac{C(3+Cr^2)(1+9\,Cr^2)}{7\pi(1+Cr^2)^3}\,.\end{align}
The first integration constant $C$ is obtain by means of the continuity of the second fundamental form, analogous condition to \eqref{FundamentalForm2Pressure}. Imposing the annulment of the latter effective radial pressure at the surface $\Sigma$, we get
	\ben\label{AnConstantCMPMD}
	CR^2=\frac{-7\left[1-\alpha+\beta(2-\alpha)\right]+\sqrt{\left[57(1-\alpha)+\beta(169-57\alpha)\right](1-\alpha)(1+\beta)}}{2\,(1-\alpha)+\beta(9-2\alpha)}\,.\een
Both the integration constant $C$ as well as the effective pressure $\bar{p}_r$ recover the corresponding values: \eqref{AnConstantCMP} and \eqref{AnRadialPressureMP} in the limit of $\beta\rightarrow0$ or \eqref{AnConstantCMD} and \eqref{AnRadialPressureMD} when $\alpha\rightarrow0$. Besides, this constant is required to plot the thermodynamic parameters. 
	\begin{figure*}[b!]
	\centering
	\includegraphics[width=.325\linewidth]{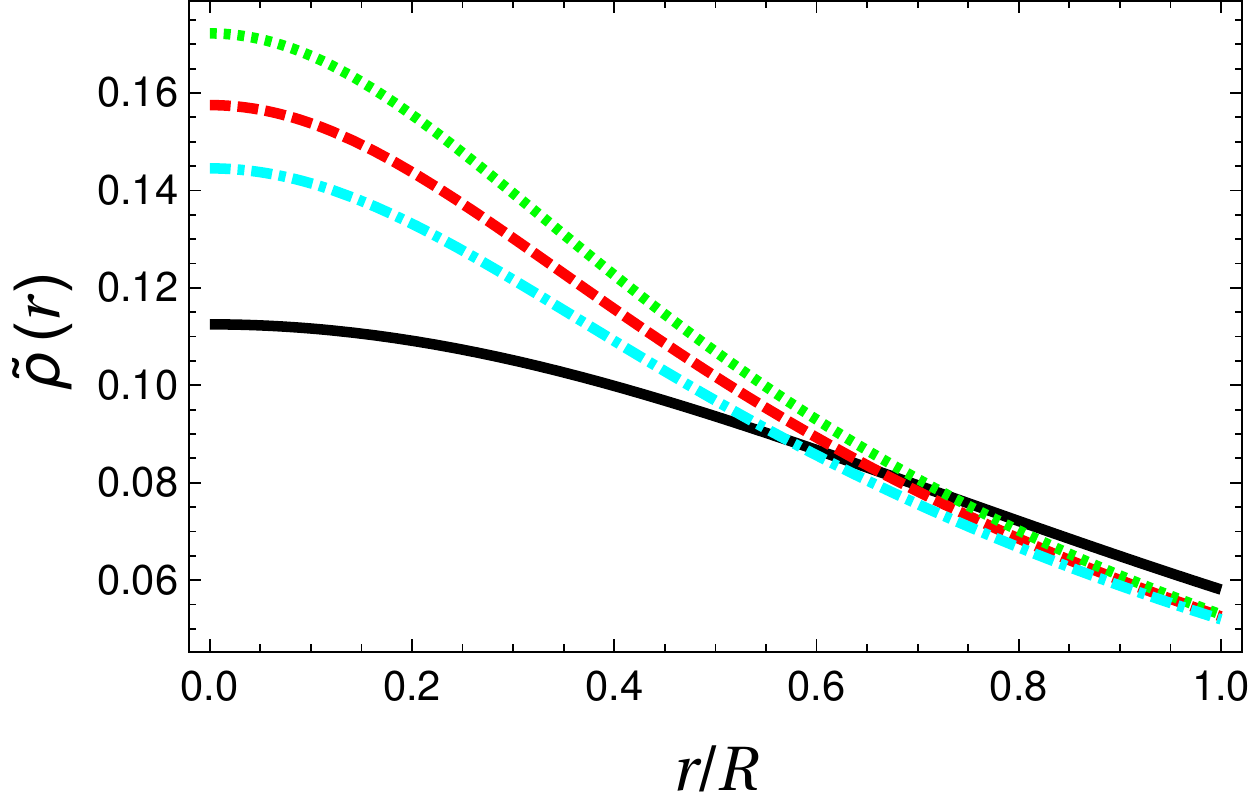}
	\includegraphics[width=.325\linewidth]{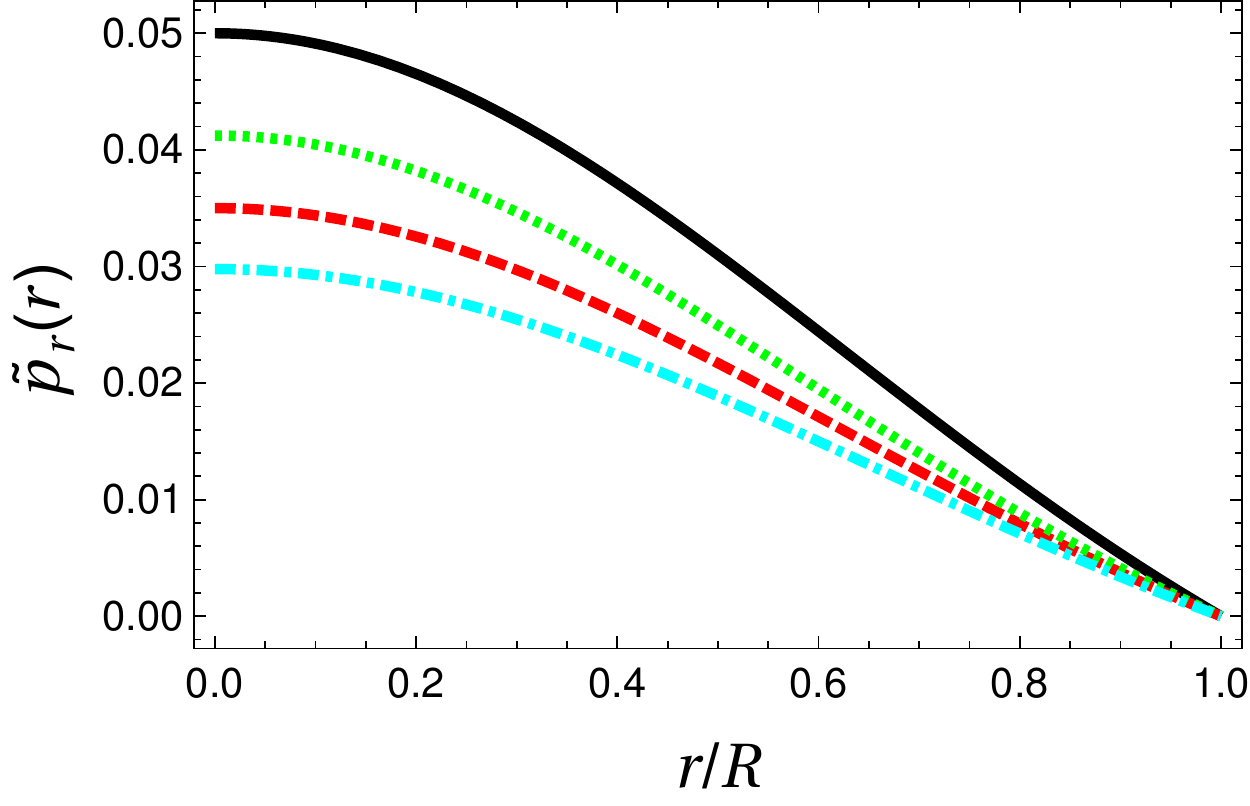}
	\includegraphics[width=.325\linewidth]{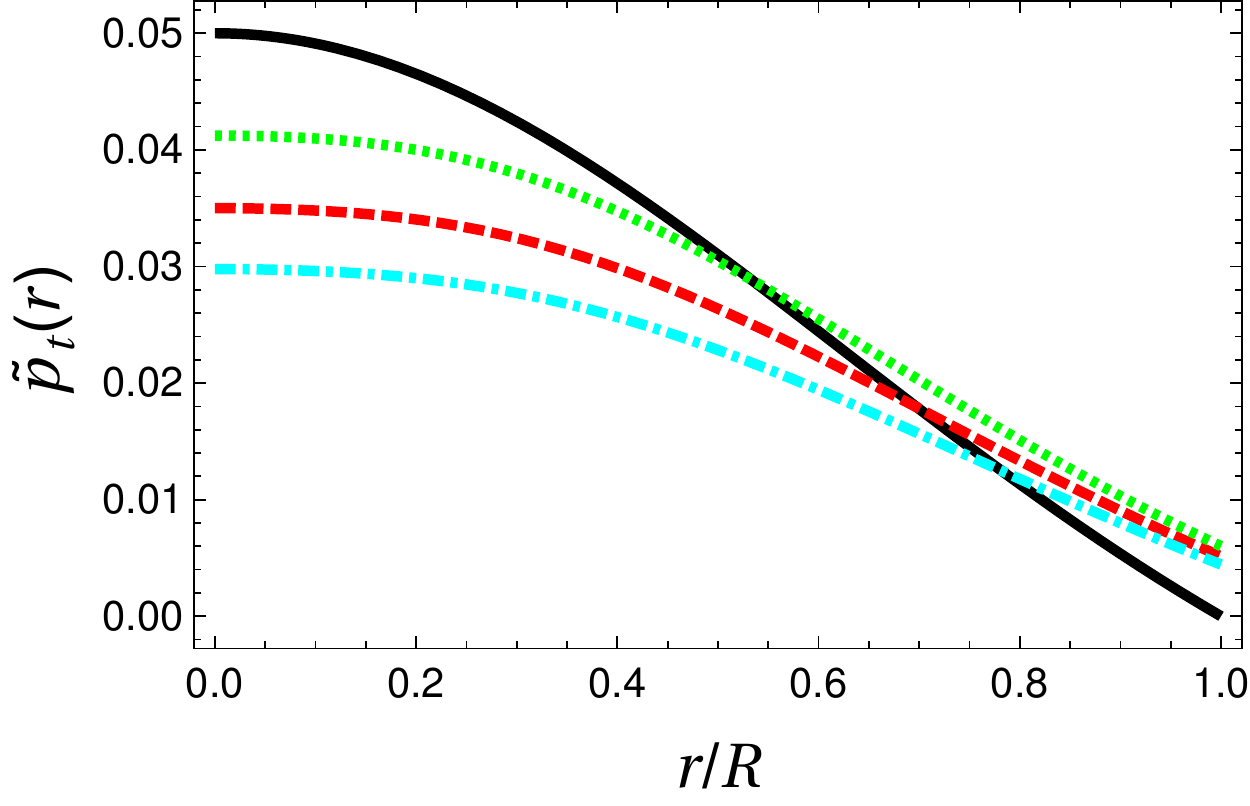}
	\vspace{30pt}
	\caption{Effective thermodynamic quantities for different values of the parameters $\{\alpha,\beta\}$. The solid black line represents the standard Durgapal--Fuloria solution ($\alpha=\beta=0$). The dashed red line is the anisotropic solution used as a seed $\{\alpha=0.3,\beta=0\}$. This configuration is {\it minimally deformed} by a $\psi$--sector. Two different solutions are presented: dotted green line for $\{\alpha=0.3,\beta=0.03\}$ and the dotted-dashed cyan line for $\{\alpha=0.3,\beta=-0.03\}$.}\label{PlotMPMD}\end{figure*}

In Figure \ref{PlotMPMD} we present the corresponding evolution of the parameters of the theory: for a comparison, we include also the Durgapal--Fuloria isotropic solution ($\alpha=\beta=0$ in solid line).
If for instance, one of the couplings move away from zero but the other remains null, the thermodynamic quantities behave as in Figures \ref{PlotMP} (if $\alpha\neq0$) or \ref{PlotMD}  (if $\beta\neq0$), as it is expected. After this, we plot the anisotropic seed to be {\it minimally deformed} by fixing the coupling $\alpha$ (red dashed-line). Finally, $\beta$ drifts away the parameters again. We choose one smaller order or magnitude for the second deformation to make notorious the effect over the seed solution. An important statement is that as the seed is anisotropic and the corresponding tangential pressure is nonnull over the surface, then there is no restriction for $\beta$ to be negative. $\beta$ is allowed to be positive until either it decrease the tangential pressure until it becomes null, or the anisotropy becomes unstable.

\begin{figure}[b!]
	\centering
	\includegraphics[width=.55\linewidth]{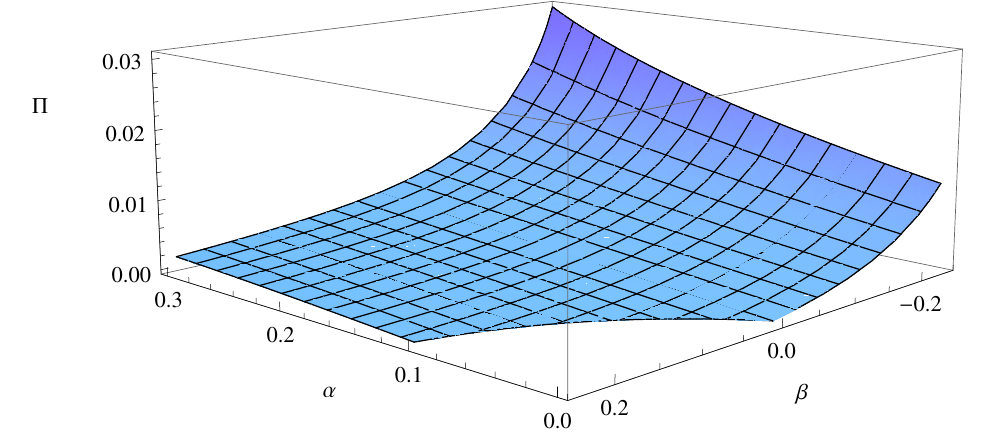}\vspace{30pt}
	\caption{\label{PlotAni} 
	The figure illustrate the anisotropy $\Pi$ as a function of the couplings $\{\alpha, \beta\}$ evaluated at the star's surface $\Sigma$. The region where $\alpha<0.1$ and $\beta>0$, the anisotropy is unphysical.}\end{figure}

Lest we forget, we include the expression of the two remaining parameters: first the effective density $\bar{\rho}=\widetilde{\rho}+\beta{\psi_t}^t$, whose structure is analogous to \eqref{AnDensityMD},
	\ben\label{AnDensityMPMF}
	\bar{\rho}(r;\alpha,\beta)=(1+\beta)\widetilde\rho\een
with $\widetilde{\rho}$ the seed density \eqref{AnDensityMP}. And secondly, the anisotropic tangential pressure $\bar{p}_t=\widetilde{p}_t-\beta{\psi_{\varphi}}^{\varphi}$ that can be written as
	\ben\label{AnTanPressureMPMF}
	\bar{p}_t(r;\alpha,\beta)= \bar{p}_r(r;\alpha,\beta)+\Pi(r;\alpha,\beta)\,.\een
The anisotropy is now written as
	\ben\label{doublePi}
	\Pi(r;\alpha,\beta)=-\beta\,\frac{C^2r^2}{\pi(1+Cr^2)^2}+(1+\beta)\alpha\,\frac{6C^2r^2(1+3Cr^2)}{\pi(1+Cr^2)(1+9Cr^2)^2}\;.\een
It is important to remark that in all expressions we recover both previous limits when either $\alpha$ or $\beta$ are set to zero, and the standard Durgapal--Fuloria solution for $\alpha=\beta=0$.
Let us conclude presenting the profile of the anisotropy over the surface $\Sigma$ of the sphere.
In Figure \ref{PlotAni}, we plot the function $\Pi$ from Eq. \eqref{doublePi} versus the couplings $\alpha$ and $\beta$.
Nearby $\alpha$ closer to zero and $\beta$ positive, the anisotropy becomes unphysical ($\Pi<0$); thus deformations with this parameters are prohibited and excluded from the physical surface.
In particular this procedure extends the range of physical values for $\beta$. Each time $\alpha$ grows, new possible values of $\beta>0$ are released.

Besides should be noted that the anisotropy can be interpreted as a lineal combination of the first two studied cases
	\ben\label{AnisotropyMPMD}
	\Pi(r;\alpha,\beta) = \Pi(r;\beta) + (1+\beta) \Pi(r;\alpha)\,;\een
Each single parameter anisotropy have been computed in \eqref{AnAnisotropyMP} and \eqref{AnAnisotropyMD}. In this case, the resulting $\Pi(\alpha,\beta)$ no longer coincides with $\beta({\psi_r}^r-{\psi_t}^t)$ as mentioned in \eqref{Anisotropy}, because the seed is no more an isotropic fluid. Immediately, the linearity is translated to the stress-energy tensor; the components of the new $\psi$--sector can be written as a combination of the single {\it minimal geometric deformations} computed in Section \ref{DFAnisotropic}
	\ben\label{PSi}
	\psi_{\mu\nu}=\theta_{\mu\nu}^{\scriptscriptstyle (density)}+\alpha\,\theta_{\mu\nu}^{\scriptscriptstyle (pressure)}\,.\een
making the `additive' character of the method manifest. If one starts with any perfect solution of GR, $T^{\scriptscriptstyle (PF)}_{\mu\nu}$, a {\it minimal deformation} induced by an anisotropy subjected to a pressure structure, makes the stress energy tensor to become $\widetilde{T}_{\mu\nu}=T^{\scriptscriptstyle (PF)}_{\mu\nu}+\alpha\,\theta_{\mu\nu}^{\scriptscriptstyle (pressure)}$. After this, a second {\it minimal deformation} acts over the already anisotropic solution, but now subjected to a density constrain. The new contribution is given by \eqref{PSi}, therefore the effective energy-momentum tensor \eqref{StressTensorEffectiveMPMD} is decomposed as
	\ben\label{StressTensorEffectiveMDMP}
	\overline{T}_{\mu\nu}=T^{\scriptscriptstyle (PF)}_{\mu\nu}+\alpha\,\theta_{\mu\nu}^{\scriptscriptstyle (pressure)}+\beta\,\bigl[\theta_{\mu\nu}^{\scriptscriptstyle (density)}+\alpha\,\theta_{\mu\nu}^{\scriptscriptstyle (pressure)}\bigr]\,.\een
This expression states the noncommutative structure of the MGD-decoupling method; the order in which the deformations take place matters. For instance, if we take as a seed the solution found in Section \ref{DFAnisotropicMD} where the deformation obeys a density--like constrain, and then we deform the anisotropic solution with a different constrain analogous to \eqref{MimicConstrainP}, the noncommutative character of the theory becomes manifest. The equations \eqref{AnisotropyMPMD} and \eqref{PSi} change their form and become
	\ben\begin{split}
	&\Pi(r;\beta,\alpha) = \Pi(r;\alpha) + (1-\alpha) \Pi(r;\beta)\,,
	\\
	&\psi_{\mu\nu}=\theta_{\mu\nu}^{\scriptscriptstyle (pressure)}-\beta\,\theta_{\mu\nu}^{\scriptscriptstyle (density)}\,;
	\end{split}\een
respectively. Likewise, the effective energy-momentum tensor \eqref{StressTensorEffectiveMDMP} becomes
	\ben
	\overline{T}_{\mu\nu}=T^{\scriptscriptstyle (PF)}_{\mu\nu}+\beta\,\theta_{\mu\nu}^{\scriptscriptstyle (density)}+\alpha\bigl[\theta_{\mu\nu}^{\scriptscriptstyle (pressure)}-\beta\,\theta_{\mu\nu}^{\scriptscriptstyle (density)}\bigr]\een
when the order of the two anistropizations is reversed.
The reason of the noncommutativity is that the coefficients involved in the linear combinations of the components of the anisotropic tensor $\theta_{\mu\nu}$  depends explicitly on the coupling constants. This is not surprising at all; we are dealing with nonlinear differential equations. The commutativity is likely to be lost in this kind of systems.

From a perturbation theory point of view, one can think that the deformations over the metric ({\it zero} order) is due to the existence of the anisotropic term which acts at ${\cal O}(\alpha)$; being $\alpha$ the coupling strength to the anisotropies.
We must emphasize that, although the MGD approach seems like a perturbation technique, the method, in fact, is not, and this is easily visualized by noticing that the couplings do not necessarily have to be small, which is a crucial ingredient in perturbation theories.
The deformation being a perturbation is just a well behaved limit of the theory, and means that we can softly deform the seed configuration.
Being the theory noncommutative, successive and mixed perturbative deformations give different configurations depending on the order in which each of them are implemented.
This provides infinite manners of deforming realistic configurations controlling rigorously the physical acceptability of the resulting anisotropic distribution.

\section{Conclusions}
In this paper we have presented different branches of solutions that models non-rotating and uncharged anisotropic superdense stars. Each anisotropic branch opens a possibility for new physically acceptable configuration obtained by guided deformations over the isotropic Durgapal--Fuloria stars, and exemplify some possible anisotropic distributions among the many that the MGD method generates.
This prescription has been design to decouple the field equations of static and spherically symmetric self-gravitating systems. 
It associate the anisotropic sector with a deformation over the geometric potentials. In this work we have reported radial deformations only, but extensions to the temporal deformation may bring intriguing results. 
After the decoupling, one obtains a sector which solution is already known (seed sector) and the anisotropic sector which obeys a set of simpler `pseudo-Einstein' equations associated to the metric deformation. It is worth to note that the {\it minimal geometric deformations} stem in an exclusively gravitational interaction between sectors; i.e. there is no exchange of energy-momentum among them.

When the equations are decoupled, no new information is introduced. Then we have an underdetermined system of equations; consistent constrains are needed. We have shown how intuitive constrain leads to new physical anisotropic solutions. Variations in the couplings between the seed and the anisotropic sector reveals consistent evolution of the thermodynamical parameter giving to the MGD method a new prove of validity.
We also have discussed the observational features of the anisotropic sectors. When the anisotropy changes the compactness of the star, the observed redshift increases as it is expected. Not all anisotropic contributions have observational effects because some anisotropies only redistribute the thermodynamical parameters in the interior. However, when the anisotropy tweaks the compactness parameter, the star suffer a redshift. Therefore, observational data would bound the parameters of the model.

After presenting two branches of solutions that provides an infinite number of physical compact stars, we have proceed to generalize the method to deform anisotropic solutions. Any solution of Einstein equation admits a minimal deformation. Different anisotropic sources have additive effect, however these effects are noncommutative. The path to the final configuration matters, and normally deformations in reversed order produces different resulting configurations.
Hence, the method provides a `fine tunning structure' that generates an enormous amount of different physically acceptable anisotropic stars.

\section*{Acknowlegements}

The author A.R. was supported by the CONICYT-PCHA/\- Doctorado Nacional/2015-21151658.
L.G. acknowledges the FPI
grant BES-2014-067939 from MINECO (Spain).
The author C.R. was supported by Conicyt PhD fellowship No 21150314.

\end{document}